%% file: ssWW_PRD.tex
\newcommand*{\ATLASLATEXPATH}{}

\pdfoutput=1 
\documentclass[atlasstyle,cernpreprint,txfonts,USenglish,texlive=2016]{\ATLASLATEXPATH atlasdoc}

\usepackage{atlaspackage} 
\usepackage{atlasbiblatex}
\usepackage{atlasphysics} 

\graphicspath{{logos/}{figures/}}

\usepackage{ssWW_defs}
\usepackage{placeins}
\usepackage{multirow}
\usepackage{afterpage}
\usepackage{threeparttable}

\AtlasTitle{Measurement of \ssww\ vector-boson scattering and limits on anomalous quartic gauge couplings with the ATLAS detector}

\PreprintIdNumber{CERN-EP-2016-167}

\AtlasJournal{Physical Review D}

\AtlasAbstract{ %
This paper presents the extended results of measurements of \sswwjj\ production and limits on anomalous quartic gauge couplings using 
20.3~fb$^{-1}$ of proton--proton collision data at $\sqrt{s}=8$ \TeV\ recorded by the ATLAS detector at the Large Hadron Collider. Events with two leptons ($e$ or $\mu$) with the same electric charge and at least two jets are analyzed. 
Production cross-sections are determined in two fiducial regions, with different sensitivities to 
the electroweak and strong production mechanisms. 
An additional fiducial region, particularly sensitive to anomalous quartic gauge coupling parameters  $\alpha_4$ and $\alpha_5$, is introduced, 
which allows more stringent limits on these parameters compared to the previous ATLAS measurement.
}

\begin{document}

\maketitle

\section{Introduction}
\label{sec:introduction}
Vector-boson scattering (VBS) processes provide a unique method to examine the mechanism of Electroweak Symmetry 
Breaking and to search for physics beyond the Standard Model (SM)~\cite{Barger, VBF2, VBF3}. 
In the SM, the Higgs boson prevents the longitudinal  scattering amplitude of the $VV \rightarrow VV$ ($V=W$ or $Z$) 
process from continuously increasing as a function of the center-of-mass energy of the diboson system, which would violate 
unitarity at energies above approximately 1 \TeV~\cite{Unitarity1, Unitarity2, Unitarity3}. 
In many new physics scenarios~\cite{BSM1, BSM2}, the Higgs boson has non-SM $HVV$ couplings below current experimental sensitivity and additional resonances are introduced to restore unitarity in the high-energy regime. 
The energy dependence of the VBS production cross-section above the Higgs boson mass scale can be used to test whether 
the Higgs boson discovered at the Large Hadron Collider (LHC)~\cite{higgs1, higgs2} unitarizes the scattering amplitude fully or only partially~\cite{VBF2}. 

\begin{figure}[htb!]
\centering
\includegraphics[width=.85\textwidth]{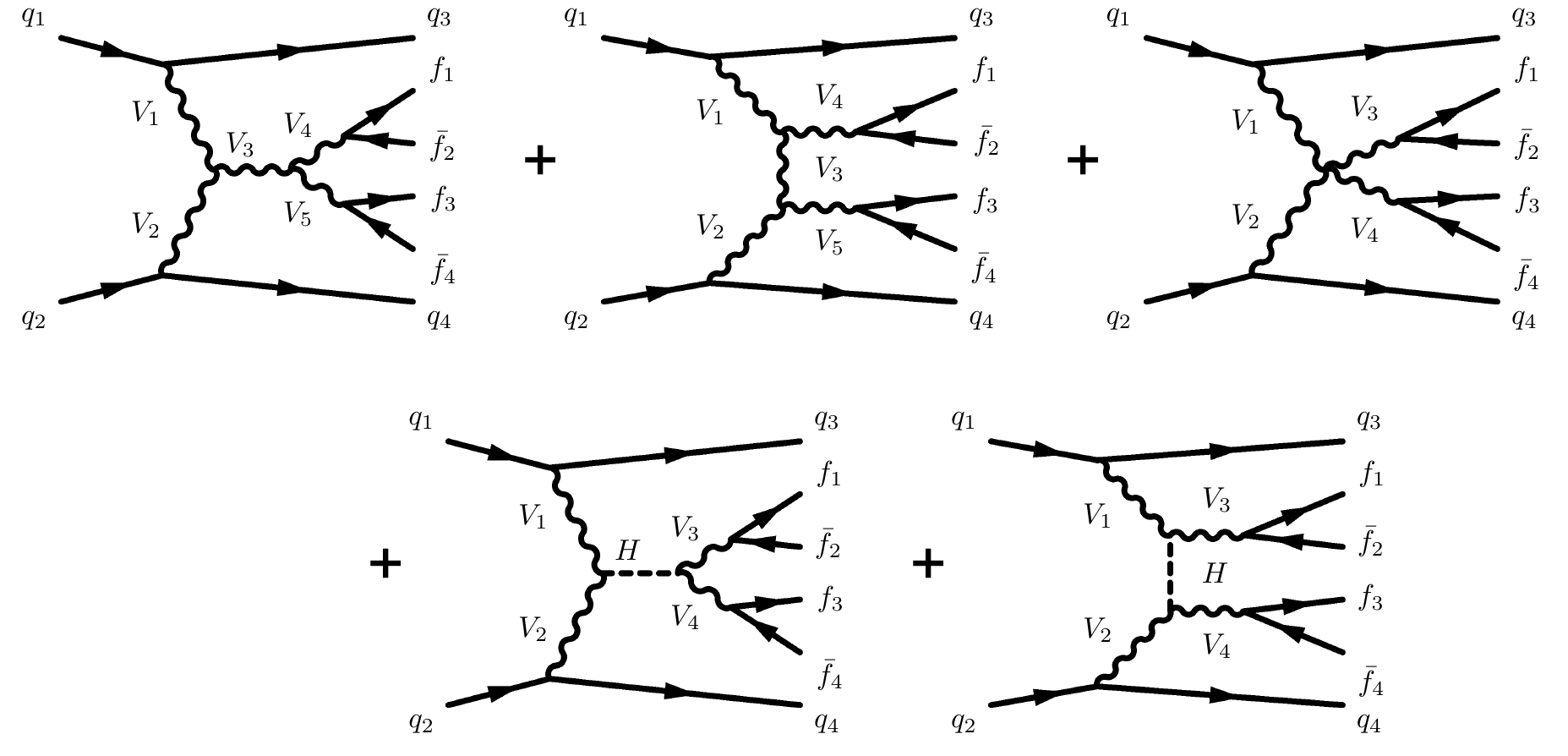}\\
\caption{Representative Feynman diagrams for \vvew production with a scattering topology including either a triple gauge boson vertex with production of a $W/Z$ boson in the $s$-channel (top left diagram), the $t$-channel exchange (top middle diagram), quartic gauge boson vertex (top right diagram), or the exchange of a Higgs boson in the $s$-channel (bottom left diagram) and $t$-channel (bottom right diagram). The lines are labeled by quarks ($q$), vector bosons ($V=W$, $Z$), and fermions ($f$).}
\label{fig:diagrams_ewk_vbs}
\end{figure}

\begin{figure}[htb!]
\centering
\includegraphics[width=.52\textwidth]{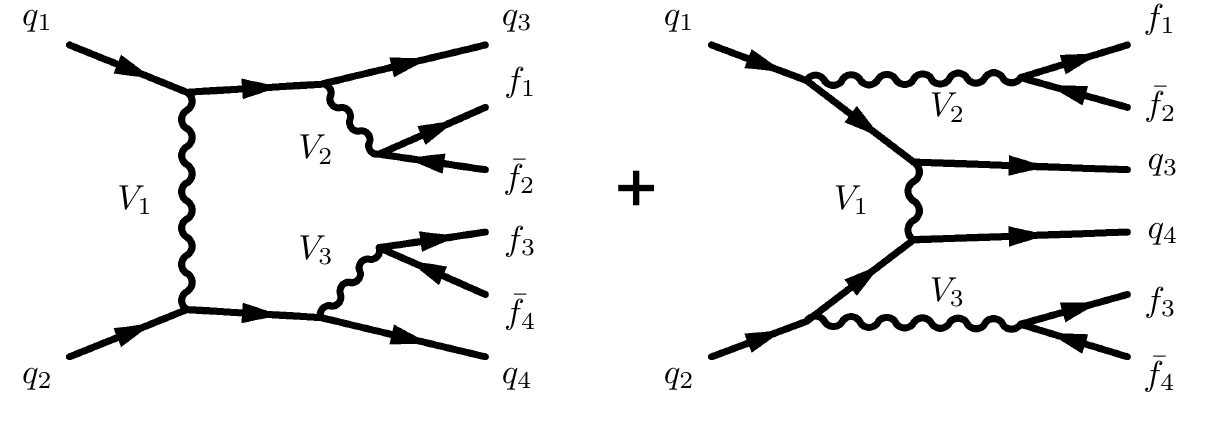}\\
\caption{Representative Feynman diagrams for \vvew production without vector-boson scattering topology. The lines are labeled by quarks ($q$), vector bosons ($V=W$, $Z$), and fermions ($f$).}
\label{fig:diagrams_ewk_nonvbs}
\end{figure}

\begin{figure}[htb!]
\centering
\includegraphics[width=.95\textwidth]{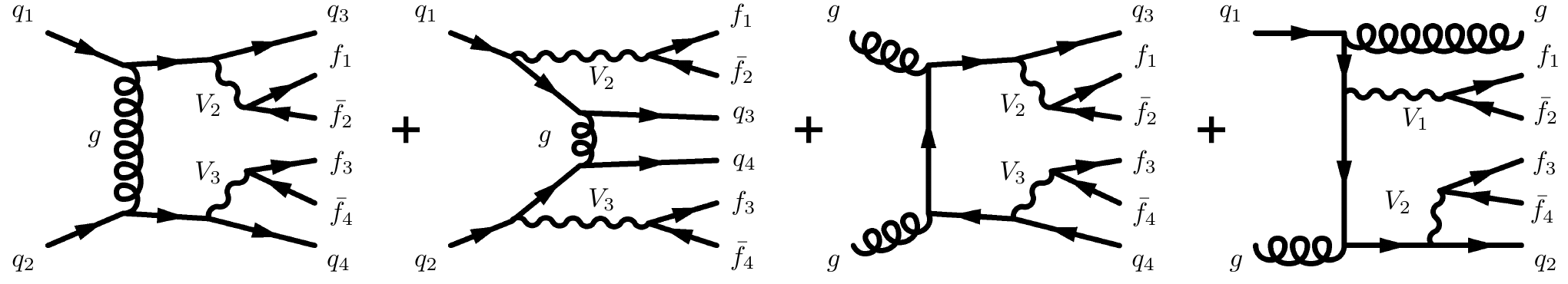}\\
\caption{Representative Feynman diagrams for \vvqcd production defined by VBS topologies with strong interaction vertices. The lines are labeled by quarks ($q$), vector bosons ($V=W$, $Z$), fermions ($f$), and gluons (g).}
\label{fig:diagrams_qcd}
\end{figure}

The VBS topology consists of a proton$-$proton collision with two initial quarks 
that each radiate an electroweak boson. The two bosons subsequently scatter and then decay. The two outgoing quarks are often close to the beam direction. 
Multiple processes can produce the same final state of two bosons ($V$) and two jets ($j$) from the fragmentation of the two outgoing quarks ($VVjj$).  
The production of $VVjj$ at tree level is composed of electroweak production involving only electroweak-interaction vertices (denoted by ``\vvew''), 
and strong production involving at least one strong-interaction vertex (denoted by ``\vvqcd'').  
The electroweak production is further categorized into two components. The first component is the EW VBS production with actual scattering of the two 
electroweak bosons. The scattering occurs via triple or quartic gauge vertices, the $s$- and $t$-channel exchange of a Higgs boson, or 
a $W/Z$ boson (throughout this paper, the notation ``$Z$ boson'' means ``$Z/\gamma^*$ boson'', unless specified otherwise). The second component is 
the EW non-VBS production with electroweak vertices only, where the two bosons do not scatter. The EW non-VBS component 
cannot be separated from the EW VBS component in a gauge invariant way~\cite{Barger}. It is therefore included in the signal generation and cannot 
be distinguished from the EW VBS. 
Representative Feynman diagrams at tree level are shown in Figure~\ref{fig:diagrams_ewk_vbs} for EW VBS production, 
in Figure~\ref{fig:diagrams_ewk_nonvbs} for EW non-VBS production, and in Figure~\ref{fig:diagrams_qcd} for \vvqcd production.
Triboson production with one of the bosons decaying hadronically also yields the same $VVjj$ final state. 
The resonant decay of a boson into two quarks can be suppressed by applying 
a requirement on the invariant mass of the two quarks. As a consequence, triboson processes are suppressed in the EW VBS signal region.

The scattering of two massive vector bosons can lead to \sswwjj, $W^+W^-jj$, $W^{\pm}Zjj$ or $ZZjj$ diboson states.
The \sswwjj\ electroweak production does not involve diagrams with the $s$-channel exchange of a Higgs boson or a vector boson, and 
the contributions from strong production are greatly suppressed due to the lack of Feynman diagrams with two gluons or one quark and one gluon in the initial state~\cite{Accomando}. 
The \sswwjj\ channel is found to have the largest cross-section ratio of electroweak to strong production~\cite{zhu}. 
Leptonic decays of the $W$ bosons ($W \rightarrow \ell \nu$)\footnote{Throughout this paper, $\ell=e, \mu$ where the notation ``electrons'' is used to mean ``electrons or positrons'' and the notation ``muons'' is used to mean ``muons or antimuons'', unless specified otherwise. Additionally, $\nu$ indicates either a neutrino or an anti-neutrino.}
are used, which allow the identification of the electric charges of the two $W$ bosons. 
The presence of two leptons with the same electric charge in the final state significantly reduces SM backgrounds. 
For these reasons, \sswwjj\ production is one of the best channels for VBS studies at the LHC~\cite{Szleper}. 

Due to the non-Abelian nature of the SM electroweak theory, gauge bosons interact with each other. 
Besides the triple $WWZ$ and $WW\gamma$ gauge boson vertices, the SM also predicts the existence of 
quartic $WWWW$, $WW\gamma\gamma$, $WWZZ$, and $WWZ\gamma$ vertices. 
Possible physics beyond the SM can affect these vertices and introduce anomalous triple gauge couplings (aTGCs) 
or anomalous quartic gauge couplings (aQGCs). 
An effective field theory (EFT) framework~\cite{Appelquist, Longhitano, Kilian, kmatrix} provides a generic platform for introducing the effect of new physics 
by adding additional terms in the SM chiral Lagrangian. 
The lowest-order terms contributing to aQGCs are the dimension-four operators $\LL_4$ and $\LL_5$: 

\begin{equation}
\alpha_4 \LL_4 = \alpha_4 \left[ \text{tr} (V_\mu V_\nu ) \right]^2 \;\; \mbox{and}\;\; \alpha_5 \LL_5 = \alpha_5 \left[ \text{tr} (V_\mu V^\mu) \right]^2,
\end{equation}

where $\alpha_4$ and $\alpha_5$ are dimensionless anomalous coupling parameters and 
$V_{\mu}=\Sigma (D_\mu \Sigma)^\dagger$ with $D_\mu$ being the covariant derivative operator. The field $\Sigma$ is a $2 \times 2$ matrix, which 
transforms  as $\Sigma \rightarrow U \Sigma V^\dagger$ under local SU(2)$_{\textrm L}$ transformations $U$ and U(1)$_{\textrm Y}$ transformations $V$. 

The EFT approach is applicable to many models of physics beyond the SM including, but not limited to, two- or multi-Higgs-doublet models, extended scalar sectors, Technicolor models, models of complete or partial compositeness, Little Higgs models, Twin Higgs models, etc. 
For example, certain heavy resonances would manifest as nonzero values of the $\alpha_5$ coupling parameter among others, but not influence $\alpha_4$~\cite{Baak:2013fwa}. 
While other models of physics beyond the SM such as a Higgs triplet, $W'/Z'$, or Kaluza--Klein graviton would manifest as nonzero parameter points in the $(\alpha_4,\alpha_5)$ plane~\cite{Reuter:2013gla}. 

Searches for processes containing QGCs have been performed by previous experiments, for example, 
$e^+ e^- \rightarrow WW\gamma, \nu\nu\gamma\gamma, qq\gamma\gamma$~\cite{lep-searches1,lep-searches2,lep-searches3,lep-searches4} 
by the LEP experiments, $p\bar{p} \rightarrow pW^+W^-\bar{p} \rightarrow p e^+ \nu e^- \bar{\nu} \bar{p}$ 
by the D0 experiment~\cite{d0_wwgammagamma}, $pp \rightarrow WV\gamma \rightarrow \ell\nu qq\gamma$~\cite{cms-wvgamma} and 
$pp \rightarrow pW^+W^-p \rightarrow pe^{\pm}\nu\mu^{\mp}\nu p$~\cite{cms-ppww-new} by the CMS experiment, 
$pp(\gamma\gamma) \rightarrow pW^+W^-p \rightarrow pe^{\pm}\nu\mu^{\mp}\nu p$~\cite{atlas_ppww} and $pp \rightarrow pW\gamma \gamma p \rightarrow p \ell \nu \gamma \gamma p$~\cite{atlas_wgammagamma} by the ATLAS experiment. 
None of these processes have been observed above 5 sigma significance, which is expected due to their low SM cross sections and large backgrounds. These results are used to set limits on corresponding aQGCs with at least one photon involved.

Experimental investigation of QGCs with four massive vector bosons has only been attempted at the LHC. Using 20.3 fb$^{-1}$ of data 
collected at $\sqrt{s}=8$ \TeV, evidence of \ssww\ decaying to $\ell^\pm \nu \ell^\pm \nu$ in association with two jets was 
recently presented~\cite{ssWWPRL} 
by the ATLAS Collaboration. 
Similar results were obtained by the CMS Collaboration~\cite{ssWWCMS} in the same final state. 
ATLAS has published a search for $WZ$ production in association with two
jets~\cite{atlas_wz}, $WW/WZ$ production in association with a high-mass dijet
system~\cite{WW-WZ}, and $WWW$ production~\cite{WWW}.  This paper completes and
extends the results presented in the form of a letter in Ref.~\cite{ssWWPRL}. An
updated Monte Carlo simulation for the signal is used, and a new signal region
more sensitive to aQGCs is developed and more stringent limits on $\alpha_4$ and
$\alpha_5$ are derived.


\section{The ATLAS detector}
The ATLAS detector~\cite{Atlas_detector} is a multipurpose particle detector designed to measure a wide range of physics processes from $pp$ collisions at the \TeV\ scale. 
It consists of an inner tracking detector (ID), calorimeters, a muon spectrometer (MS), and solenoidal and toroidal magnets in a cylindrical geometry 
with forward-backward symmetry.\footnote{The ATLAS reference system is a Cartesian right-handed coordinate system with its origin at the nominal interaction point (IP) in the 
center of the detector and the $z$-axis along the beam direction. The $x$-axis points from the IP to the center of the LHC ring and the $y$-axis points upward. 
Cylindrical coordinates ($r,\phi$) are used in the plane that is transverse to the beam direction, where $\phi$ describes the azimuthal angle 
around the beam pipe as measured from the positive $x$-axis. 
Rapidity ($y$) is defined as $y=1/2 \times \ln [(E+p_z)/(E-p_z)]$, where $E$ ($p_z$) is the energy (the $z$-component of the momentum) of a particle. 
Pseudorapidity ($\eta$) is defined as $\eta = -\ln(\tan\theta/2)$ where $\theta$ is 
the polar angle. 
Transverse momentum ($\pt$) is defined relative to the beam axis and is calculated as $\pt=p \sin \theta$ where $p$ is the momentum. 
The distance between two objects in the $\eta$--$\phi$ space is defined as $\Delta R=\sqrt{(\eta_1-\eta_2)^2 + (\phi_1-\phi_2)^2}$ where $\eta_{1,2}$ ($\phi_{1,2}$) 
represents the pseudorapidities (azimuthal angle) of the two objects.}

The ID consists of three subdetectors. The pixel detector and semiconductor tracker (SCT) are composed of silicon pixel and microstrip 
detectors and extend to $|\eta|$~$=$~2.5. In this region, the pixel detector has 3 cylindrical layers and the SCT has 4 layers. The transition radiation tracker (TRT) is built of gas-filled 
straw-tube detectors and extends to $|\eta|$~$=$~2.0. The ID is surrounded by a thin superconducting solenoid magnet that 
creates a 2~T axial magnetic field for charged-particle momentum measurements.

The calorimeter system consists of electromagnetic (EM) and hadronic calorimeters. 
A high-granularity sampling calorimeter with lead absorber layers and liquid argon 
(LAr) measures the energy and position of electromagnetic showers in the pseudorapidity region of $|\eta|$~$<$~3.2. Hadronic showers are measured 
by steel and scintillator tile calorimeters for $|\eta|$~$<$~1.7 and copper/LAr calorimeters for 1.5~$<$~$|\eta|$~$<$~3.2. The forward calorimeter 
extends the coverage, spanning 3.1~$<$~$|\eta|$~$<$~4.9 with additional copper/LAr and tungsten/LAr calorimeters. 

The MS covers the pseudorapidity range of $|\eta|<2.7$ and is instrumented with separate trigger and precision tracking chambers. 
A precision measurement of the track coordinates in the bending direction of the toroidal magnetic field is provided by drift tubes up to $|\eta|=2.0$. 
At larger pseudorapidities, cathode strip chambers with higher granularity are used in the innermost station covering $2.0<|\eta|<2.7$. The 
muon trigger system consists of resistive plate chambers in the barrel ($|\eta|<1.05$) and thin gap chambers in the endcap regions ($1.05<|\eta|<2.4$). 

A three-level trigger system is used to record the events used in this analysis. The level-1 trigger is implemented in hardware and reduces the event 
rate to about 75 kHz. This is followed by two software-based trigger levels that together 
reduced the event rate to about 600 Hz during the 2012 data-taking period.

\section{Event selection}
\label{sec:Analysis_Selection}
Candidate events are collected by single-lepton triggers with thresholds of $\pt = 36$ \GeV\ (muons) or $\pt = 60$ \GeV\ (electrons)
or single-isolated-lepton triggers with a lower threshold of $\pt = 24$ \GeV .
The events must also occur during stable beam conditions and with the relevant detector systems functional.
The resulting total integrated luminosity is 20.3 fb$^{-1}$ with an uncertainty of 2.8\%~\cite{lumi}.

Tracks used in this analysis are reconstructed using an ``inside-out'' algorithm starting with seeds made from hits in the 
pixel detector and the first layer of the SCT and attempting to extend these into the remaining silicon layers and 
finally into the TRT~\cite{trackReco}. Proton$-$proton interaction vertices are reconstructed by 
extrapolating the $z$-position of tracks at the beamline, grouping two or more tracks into vertex candidates,
and then reconstructing the vertex position and its corresponding error matrix.
Tracks incompatible with the vertex by more than seven standard deviations are used to look for 
additional vertices. The vertex with the largest sum of squared transverse momenta of associated tracks ($\sum \pt^{2}$) 
is taken to be the primary vertex. The primary vertex is required to have at least three associated tracks with $\pt>0.4$ \GeV.

Three types of lepton identification criteria are defined for signal selection and background rejection, which are non-exclusive: 
a tight lepton criterion used to select the final two same-electric-charge leptons, 
a veto lepton used to reject events with an additional lepton present in $W^{\pm}Z$ or $ZZ$ events, and a loose lepton category
used to estimate the background contribution from events with non-prompt leptons from in-flight hadron decays or with jets misidentified as leptons. 

Electrons are reconstructed from a combination of track information in the ID and cluster information in the electromagnetic calorimeter. 
Tight electrons must satisfy identification criteria similar to the tight definition used in Refs.~\cite{eid1,eid2, eid3}, which includes 
requirements on the electron track, the shape of the shower in the EM calorimeter, and the ratio of energies deposited 
in the EM and hadronic calorimeters. 
Additionally, the track hit information is used to identify and remove electrons arising from photon conversions. Electron candidates must have $\pt>25$ \GeV\ 
and $|\eta|<2.47$. Electrons within the transition region ($1.37<|\eta|<1.52$) between the EM barrel and endcap calorimeters are excluded. The 
transverse ($d_0$) and longitudinal ($z_0$) impact parameters must satisfy $|d_{0}/\sigma_{d_0}|<3$ and $|z_{0} \times \sin\theta|<0.5$ mm, 
where $\sigma_{d_0}$ is the uncertainty in the measurement of $d_0$. Finally, calorimeter and tracking isolation selections are applied as 
follows: the sum of the transverse energies of all calorimeter clusters ($E_{\textrm T}^{\textrm iso}$) and the sum of the transverse momenta of tracks 
($p_{\textrm T}^{\textrm iso}$) within a cone of size $\Delta R =0.3$, 
are required to be less than 14\% and 6\% of the electron's transverse energy, respectively. 
The energy from the electron itself is excluded in the calculations of $E_{\textrm T}^{\textrm iso}$  and $p_{\textrm T}^{\textrm iso}$. 

Veto and loose electrons are only required to pass a loose identification selection defined in Ref.~\cite{eid1}. 
The $\pt$ threshold is lowered to 7 \GeV, and the tracking isolation requirement is removed for veto electrons. For loose electrons, the impact 
parameter requirements are loosened to $|d_{0}/\sigma_{d_0}|<10$ and $|z_{0}\times \sin \theta|~<~5$~mm, and the calorimeter and tracking 
isolation criteria are $0.14<E_{\textrm T}^{\textrm iso}/\pt<2$ and $0.06<p_{\textrm T}^{\textrm iso}/\pt<2$. 

Muons are reconstructed from tracks in the ID and MS and fall into one of three categories: combined, standalone, and tagged~\cite{muonID}. 
Combined muons contain matching tracks in the ID and MS. Standalone muons consist only of a track in the MS, while tagged 
muons have an ID track that is matched to a track segment in the MS. In this analysis, tight muons are required to be reconstructed 
as combined muons with the same electric charge measured in the ID and MS. They must have $p_{\textrm T}~>~$25 \GeV\ and $|\eta|~<~$2.5. 
The ID tracks associated with these muons must pass a number of quality requirements. The number of hits or dead sensors crossed in 
the pixel detector must be at least one, and in the SCT this number must be at least five. For muons with $0.1<|\eta|<1.9$, 
the track must have at least six hits in the TRT, and the fraction of these that are outliers must not exceed 90\%. Tight muons have 
the same impact parameter requirements as tight electrons and have calorimeter and tracking isolation requirements 
defined by $E_{\textrm T}^{\textrm iso}/p_{\textrm T}<0.07$ and $p_{\textrm T}^{\textrm iso}/p_{\textrm T}<0.07$ where a cone of size $\Delta R = 0.3$ is used.

The selection of veto muons includes standalone and tagged muons. The $\pt$ threshold is lowered to 6~\GeV, the calorimeter isolation 
requirement is dropped, and the track isolation selection is loosened to be less than 15\% of the muon $\pt$. Loose muons must be 
combined muons, but just as for loose electrons, the impact parameter requirements are loosened to 
$|d_{0}/\sigma_{d_0}|<10$ and $|z_{0}\times \sin \theta|<5$ mm, and the calorimeter and tracking 
isolation criteria are $0.07<p_{\textrm T}^{\textrm iso}/\pt<2$ and $0.07<p_{\textrm T}^{\textrm iso}/\pt<2$. 

To improve agreement between data and simulation, lepton selection efficiencies are measured in both
data and simulation, and correction factors are applied to the simulation to account for differences with respect
to data~\cite{eid3, muonID}. Furthermore, the simulation is tuned to reproduce the calorimeter energy and the muon momentum
scales and resolutions observed in data. The simulation also includes modeling of additional $pp$ interactions in the same and neighboring bunch crossings.

Jets are reconstructed from topological clusters in the calorimeter using the anti-$k_{t}$ algorithm~\cite{antikt} with a radius parameter of 0.4~\cite{jetPerf}. 
Jets are required to have $\pt>30$ \GeV\ and $|\eta|<4.5$. In order to 
reduce the probability of selecting a jet from a pileup interaction, jets with $|\eta|<2.4$ and $p_{\textrm T}<50$ \GeV\ 
are required to have a jet vertex fraction greater than 50\%. The jet vertex fraction is defined as the ratio of the sum of the \pt\ of all tracks associated with both the jet and the primary vertex to the sum of the \pt\ of all tracks in the jet~\cite{jvf}. 
Jets stemming from the fragmentation of a charm or bottom quark are identified with a neural network discriminator using 
input variables related to the impact parameter significance of tracks in the jet and secondary vertices reconstructed from these 
tracks~\cite{bjetPerf}. 
The jet is classified as a $b$-jet if the output of this neural network discriminator exceeds a working point chosen to 
have a 70$\%$ efficiency for identifying jets from top quarks containing $b$-hadrons.

The measurement of the two-dimensional missing transverse momentum vector $\vec{E}_{\textrm T}^{\textrm miss}$ and its magnitude $E_{\textrm T}^{\textrm miss}$~\cite{met} 
is based on the measurement of all topological clusters in the calorimeter, and muon tracks reconstructed by the ID and MS. 
The energies of clusters in the calorimeter are calibrated according to their association with a reconstructed object. 

In order to deal with the case where a single particle is reconstructed as more than one object, an overlap removal procedure is followed. 
If the event contains a tight electron and a jet with $\Delta R(e, j)<0.3$, the jet is removed since it is likely that it corresponds to the electron energy deposits 
picked up by the jet reconstruction algorithm. If the same is true for a jet and a tight 
muon, the event is rejected since the muon likely originates from the decay of a hadron within the jet. When estimating the background 
from non-prompt leptons, jets are also removed if they fall within $\Delta R = 0.3$ of a loose lepton. 
For electrons and muons seperated by $\Delta R<0.1$, the electron is removed since it is likely that it originates from 
a photon radiated from the muon. 

Signal candidate events are selected by requiring two tight leptons with the same electric charge and an invariant mass 
($m_{\ell\ell}$) greater than 20 \GeV. 
Three final states are considered based on the lepton flavor, namely \ssee, \ssem, and \ssmm.
To reduce background contributions from the $W^{\pm}Z$ and $ZZ$ processes, events with a third lepton of the veto type are rejected. 
An additional requirement is made in the \ssee\ final state that the invariant mass of the two electrons differs from the combined world average of the $Z$ pole mass~\cite{pdg} by 
at least 10 \GeV. This selection criterion reduces the background from the $Z (\rightarrow e^+ e^-)+$jets process where one electron's charge is misidentified. 
Since two neutrinos are produced from the decays of the two $W$ bosons, $E_{\textrm T}^{\textrm miss}$ is required to be 
greater than 40 \GeV. Events are required to have at least two jets. In order to reduce the background from top-quark pair and single top-quark production, 
the event is rejected if any jet is classified as a $b$-jet. Remaining events with an invariant mass of the two leading-$\pt$ jets ($m_{jj}$) 
greater than 500 \GeV\ are selected. This selection level defines the inclusive signal region (denoted by ``Inclusive SR''), and both the 
electroweak and strong production of $W^{\pm}W^{\pm}jj$ are treated as signal. 
The VBS signal region (denoted by ``VBS SR'') is defined to consist of events in 
the inclusive signal region for which the separation in rapidity between the two leading-$\pt$ jets 
($|\Delta y_{jj}|$) is greater than 2.4. In this region only the electroweak production is considered as signal. 
The third signal region (denoted by ``aQGC SR'') additionally requires the estimated transverse mass of the 
\WW~system to be greater than 400 \GeV\ in order to optimize the sensitivity to the new-physics parameters $\alpha_4$ and $\alpha_5$. 
The variable, $m_{WW,\mathrm{T}}$, is defined as

\begin{equation}
m_{WW,\mathrm{T}} = \sqrt{\left(\mathbf{P}_{\ell_1} + \mathbf{P}_{\ell_2} + \mathbf{P}_{\met} \right)^2}
\end{equation}

where $\mathbf{P}_{\ell_1, \ell_2}$ are the four-momenta of the two selected lepton
candidates and $\mathbf{P}_{\met}$ is the massless four-vector constructed from the $\vec{E}_{\textrm T}^{\textrm miss}$ measurement with 
the $z$-component of $\mathbf{P}_{\met}$ defined as zero. 
In the aQGC SR, both the electroweak and strong production predicted by the SM are considered as background, and only the contributions due to aQGCs are considered as signal. 

Table~\ref{tab:cuts} summarizes the kinematic selection criteria used for the three signal regions.

\begin{table*}
\begin{center}
\begin{tabular}{l|l|c}   \hline
Signal Region & & \multicolumn{1}{c}{Selection Criteria}  \\  \hline \hline
\multirow{8}{*}{Inclusive} & Lepton & \multicolumn{1}{c}{Exactly two tight same-electric-charge leptons with $\pt>25$ \GeV} \\ \cline{2-3}
& Jet        & \multicolumn{1}{c}{At least two jets with $\pt>30$ \GeV\ and $|\eta|<4.5$} \\  \cline{2-3}
& $m_{\ell\ell}$        & \multicolumn{1}{c}{$m_{\ell\ell}>20$ \GeV}  \\ \cline{2-3}
& \met        & \multicolumn{1}{c}{$\met>40$ \GeV}  \\ \cline{2-3}
& $Z$ veto              & \multicolumn{1}{c}{$|m_{\ell\ell}-m_Z|>10$ \GeV\ (only for the \ssee\ channel)}  \\ \cline{2-3}
& Third-lepton veto                      & \multicolumn{1}{c}{No third veto-lepton} \\ \cline{2-3}
& $b$-jet veto                              & \multicolumn{1}{c}{No identified $b$-jets with $\pt>30$ \GeV\ and $|\eta|<2.5$} \\ \cline{2-3}
& $m_{jj}$        & \multicolumn{1}{c}{$m_{jj}>500$ \GeV}  \\ \hline
VBS & $\Delta y_{jj}$        & \multicolumn{1}{c}{$|\Delta y_{jj}|>2.4$}  \\ \cline{1-3}
aQGC & $m_{WW,\mathrm{T}}$        & \multicolumn{1}{c}{$m_{WW,\mathrm{T}}>400$ \GeV}  \\
\hline
\end{tabular}
\end{center}
\caption{Kinematic selection criteria used for three signal regions. These selection criteria are applied successively for each signal region such that the aQGC signal region has all requirements applied.}
\label{tab:cuts}
\end{table*}

\section{Monte Carlo simulation and theoretical predictions}
\label{sec:predxsec}
Monte Carlo (MC) events are simulated at $\sqrt{s}=8$ \TeV{} and processed through the full ATLAS detector simulation~\cite{atlasSim} based on {\textsc geant4}~\cite{geant}. Additional proton$-$proton interactions modeled by {\textsc pythia} 8~\cite{pythia8a, pythia8b} are included and reweighted to reproduce the observed distribution of the average number of 
proton$-$proton interactions per event. Contributions from interactions in nearby bunch crossings are also considered in the MC simulations. 
Events generated in the Inclusive and VBS signal regions are used to measure the production cross-sections, provide normalization factors for MC samples, and to compare with theoretical predictions. 
This section concentrates on the theoretical cross-sections and uncertainties for the \sswwjj-EW and \sswwjj-QCD processes in these two regions.

\bigskip 
\noindent\textbf{Definition of Inclusive and VBS fiducial phase-space regions at the particle level} \\
Two fiducial phase-space regions are defined at particle level by
selection criteria similar to the ``Inclusive SR'' and ``VBS SR'' described in Section~\ref{sec:Analysis_Selection}. 
Particle level jets are reconstructed by running the anti-$k_t$ algorithm
with radius parameter $R = 0.4$ on all observable final-state stable particles after parton showering and hadronization. 
The inclusive fiducial phase-space region is defined with the following criteria: exactly two charged leptons~(only considering electrons and muons) of the same electric charge, each with $\pt > 25$ \GeV\ and $|\eta| < 2.5$, 
and at least two particle level jets with $\pt > 30$ \GeV\ and $|\eta| < 4.5$. The jets are required to be separated from
leptons by $\Delta R(\ell, j) > 0.3$. The events are further required to have a dilepton invariant mass $m_{\ell\ell} > 20$ \GeV\ and $\pt^{\nu_1+\nu_2}>40$ \GeV, 
where $\pt^{\nu_1+\nu_2}$ is the magnitude of the vectorial sum of $p_{\textrm T}$ of the two particle level neutrinos. 
The lepton four-momentum includes contributions from photons within $\Delta R(\ell, \gamma) = 0.1$ of the lepton direction. 
The two leptons are also required to be separated by $\Delta R>0.3$. 
The two leading-$\pt$ jets are required to have $m_{jj}>500$ \GeV. 
An additional requirement of $|\Delta y_{jj}|>2.4$ is applied for the VBS
fiducial phase-space region. 

\bigskip 
\noindent\textbf{\sswwjj-EW and \sswwjj-QCD cross-sections and uncertainties} \\
Both electroweak and strong production of \sswwjj\ events are generated using
the {\textsc sherpa} version 1.4.5 event generator~\cite{sherpa} at leading order (LO) in QCD 
with up to three partons. Matrix-element and parton-shower matching for the two final-state jets are performed with the CKKW scheme~\cite{ckkw}.
Dynamic factorization ($\mu_\mathrm{F}$) and renormalization ($\mu_\mathrm{R}$) scales are set to be  

\begin{equation}\label{eqn:RFscales}
\mu_{\mathrm{F,R}} = \frac{1}{2} \sum_{i=1,2} { \left[ \pT(j_i) + \sqrt{ m^2(W_i) + \pT^2(W_i)} \right]},
\end{equation}

where $\pt(j_i)$ is the momentum of the $i^{\textrm th}$ leading-$p_{\textrm T}$ jet, and $m(W_i)$ and $\pt(W_i)$ are the mass and transverse momentum of the $i^{\textrm th}$ $W$ boson.
CT10 parton distribution functions (PDFs)~\cite{ct10} are used. 
 
The \sswwjj\ \sherpa\ samples are updated from those in the previous publication of the measurement of \sswwjj~\cite{ssWWPRL} 
to include a more accurate representation of the QED final-state radiation. The
impact of this effect reduces the final acceptance due to an additional 5\% loss of 
leptons in the lepton--jet overlap removal in both fiducial phase-space regions.

The {\textsc sherpa} cross-sections are scaled to account for the next-to-leading-order (NLO) cross-section predictions 
using {\textsc powheg-box}~\cite{powhegbox1, powhegbox2,powhegbox3} with {\textsc
pythia} 8 for parton shower and hadronization in the fiducial phase-space regions. 
The dynamic scales defined in Eq.~\eqref{eqn:RFscales} are used. 
Contributions from non-resonant production are included, but are highly suppressed.  
Interference effects between the electroweak and strong production are studied using separated and combined electroweak and strong-mediated samples. 
The cross-section for the combined sample minus the sum of the cross-sections of purely electroweakly-mediated and purely strongly-mediated samples gives 
the size of the interference effect. 
The interference is found to enhance the total signal production cross-section
by 10.7\% in the Inclusive phase-space region and 6.5\% in the VBS phase-space region. 

The prediction for \ew production is cross-checked using {\textsc vbfnlo}~\cite{VBFNLO1, VBFNLO2, VBFNLO3} 
and the results from the two generators are found to be consistent to within 5\%. 
This 5\% difference is taken as the generator uncertainty. 
Scale- and PDF-induced uncertainties are evaluated using {\textsc vbfnlo}.
Scale-induced uncertainties are estimated by varying separately the factorization and renormalization scales from the central values as listed in Eq.~\eqref{eqn:RFscales} 
by factors $\xi_\mathrm{F}$ and $\xi_\mathrm{R}$. The largest difference in the cross-section resulting from variations of (~$\xi_\mathrm{F},~\xi_\mathrm{R}$) where $\xi_\mathrm{F},~\xi_\mathrm{R}$ = 0.5, 1, or 2 
excluding extremum combinations (~$\xi_\mathrm{F}=0.5, ~\xi_\mathrm{R}=2$) and (~$\xi_\mathrm{F}=2, ~\xi_\mathrm{R}=0.5$) of scale variations is taken as the uncertainty.
 The PDF uncertainty is determined by adding in quadrature the CT10 eigenvector variations~\cite{ct10} and the difference of central values with respect to MSTW2008~\cite{mstw2008}.

Due to the selection criteria applied to jet transverse momenta and dijet mass, the parton shower has an effect on 
the fiducial cross-sections \cite{pb_ssww1,pb_ssww2,pb_ssww3,pb_ssww4}.
Two different parton-shower algorithms are applied to {\textsc powheg-box}\ NLO events and the difference in the signal yield is used to determine the uncertainty. 
The default algorithm relies on the \pythia\ 8 parton-shower model using the AU2 set of tuned parameters~\cite{pythiaAtlasTunes} for the underlying-event modeling. 
The second algorithm uses the \herwig~\cite{herwig} parton-shower model with \jimmy~\cite{jimmy} to model the underlying event. 

The NLO cross-sections for the \sswwjj-QCD production are also calculated using the {\textsc powheg-box} generator. 
Uncertainties due to the scale, PDF, and parton-shower model are evaluated in the same way as for the \sswwjj-EW production. 

Theoretical uncertainties in the predictions for \sswwjj-EW and \sswwjj-QCD production 
 in the Inclusive and VBS fiducial phase-space regions are detailed in Table~\ref{tab:vvjj_systematics}. 
The \sswwjj-EW (\sswwjj-QCD) production cross-section is predicted to be $1.00 \pm 0.06$ fb ($0.35 \pm 0.05$ fb) 
in the Inclusive phase-space region and $0.88 \pm 0.05$ fb ($0.098 \pm 0.018$
 fb) in the VBS phase-space region. 
The interference between \sswwjj-EW and \sswwjj-QCD production enhances the
 cross-section by $0.16 \pm 0.08$ fb in the Inclusive phase-space region 
and $0.07 \pm 0.04$ fb in the VBS phase-space region. 
Both the electroweak and strong production of \sswwjj{} and their
interference are treated as signal in the Inclusive phase-space region. The total
predicted signal cross-section in the Inclusive phase-space region is $1.52 \pm 0.11$ fb. For the VBS phase-space region, the electroweak production and the interference term
are included in the total predicted cross-section, which is determined to be $0.95 \pm 0.06$ fb. 
For the rest of the paper, \sswwjj-EW is used to indicate the combined contribution from the electroweak production and the interference effect, 
while \sswwjj-EW+QCD indicates contributions from both electroweak and strong production as well as the interference effect.

\begin{table*}
  \centering
  \begin{tabular}{l|r|r|r|r}
\hline
\hline
\multirow{2}[1]{*} {Source of uncertainty} &  \multicolumn{2}{c|}{\sswwjj-EW} &  \multicolumn{2}{c}{\sswwjj-QCD} \\ \cline{2-5}
  & Inclusive & VBS  & Inclusive & VBS  \\
 \hline
MC sample size    &  1\%   &  2\%   & 4\%    &  8\%   \\  
Showering model   &   2\%   &   4\%    & 3\%   & 7\%   \\ 
Scale &  2\% & 2\%  & 12\%  & 13\%  \\ 
PDF    &   2\%   & 3\%  & 2\%  & 2\%   \\ 
Generator        &   5\%   &  3\%   & 5\%    &  5\%   \\  \hline 
Total uncertainty & 6\% & 6\% & 14\% & 18\% \\
\hline
\hline
\end{tabular}
\caption{Summary of theoretical uncertainties for the \sswwjj-EW and \sswwjj-QCD
 production in the Inclusive and VBS fiducial phase-space regions.}
\label{tab:vvjj_systematics}
\end{table*}

\section{Backgrounds}
\label{sec:backgrounds}
SM background processes producing the signature of two same-electric-charge leptons and \met\ with at least two jets in the 
final state are grouped in three categories: prompt background, non-prompt background, and conversions. The prompt background is due to 
$WZ+$jets, $ZZ+$jets, or $t\bar{t}V$ production when one or more leptons are 
either not reconstructed or not identified while the remaining two prompt leptons have the same electric charge. 
The non-prompt background is due to processes with one or two jets mis-reconstructed as tight leptons. 
The main contributions come from $W+$jets, $t\bar{t}$, single top quark, and multijet production.
The conversion background events are mainly due to processes where two prompt electrons of opposite electric charge are produced but one radiates a photon that converts to $e^+ e^-$. 
The main contribution comes from $Z+$jets production 
where the $Z$ boson decays to $e^+ e^-$. 
The background estimation for the prompt background category is based on MC-simulated samples, 
while estimations for the other two categories are 
based on data-driven methods. The modeling of the backgrounds is checked in several control regions. 

\subsection{Prompt background}
\label{sec:prompt}

The main source of prompt background is $WZjj$ production where both bosons decay leptonically and one lepton lies 
outside of the detector acceptance or fails the lepton identification requirements. Similarly to $W^{\pm}W^{\pm}jj$, there are 
strong and electroweak production mechanisms for $WZjj$, which contribute about 75$\%$ and 15$\%$ of the 
prompt background, respectively. 
The two production mechanisms are generated using the {\textsc sherpa} event generator at LO in QCD with up to three partons and 
normalized to NLO cross-sections calculated with {\textsc vbfnlo} in each fiducial
phase-space region.  The CT10 PDF set is used. 
The normalization of the electroweak production of $WZjj$ contains a further complication. This process receives a contribution 
from the production of a top quark in association with a $Z$ boson and an additional parton ($tZj$), where the top quark further decays 
to a $W$ boson and a $b$-quark. This class of diagrams is taken into account in {\textsc sherpa} but is neglected in \textsc{vbfnlo}, 
even though it contributes almost a third of the events populating both
phase-space regions. To account for this, 
a new normalization is derived using the $b$-quark in the initial state to select for $tZj$ events. The samples are split into events that contain a $b$-quark in the initial state (using {\textsc sherpa} at LO) 
and events without an initial $b$-quark (using {\textsc vbfnlo} at NLO). The cross-section used to normalize the \textsc{sherpa} sample is given by 
$\sigma^{\textsc{vbfnlo}}_{\textrm fid}/A + \sigma^{\textsc{sherpa}}_{\textrm fid} \times f_{b}$, 
where $\sigma^{\textsc{vbfnlo}}_{\textrm fid}$ is the NLO cross-section calculated using \textsc{vbfnlo}, 
$\sigma^{\textsc{sherpa}}_{\textrm fid}$ is the sum of  LO cross-sections calculated with and without a $b$-quark in the initial state using \textsc{sherpa}, 
$A$ is the parton-level acceptance of the \textsc{sherpa} subsample without any $b$-quarks in the initial state, 
and $f_b$ is the fraction of generated events containing a $b$-quark in the initial state. 
The overall cross-section for the electroweak \wzjj\ production used for the normalization is $0.40 \pm 0.09$ fb ($0.34 \pm 0.09$ fb) in the Inclusive (VBS) SR, 
while the corresponding cross-section for the strong production is $1.04 \pm 0.17$ fb ($0.64~\pm~0.08$~fb). 

Other processes with two prompt leptons with the same electric charge in the final state include 
the $t\bar{t}V$ process, $ZZjj$ production, and multiple parton$-$parton interactions (MPI) in one proton$-$proton interaction. 
The sum of these backgrounds contributes less than 10\% of the total prompt background.
The $t\bar{t}V$ events are generated using {\textsc madgraph}~\cite{madgraph} with {\textsc pythia} 8 used 
for parton shower and hadronization. The CTEQ6L1 PDF~\cite{cteq6l} is used. The $ZZjj$ events are simulated using \textsc{sherpa} with the CT10 PDF set. 
MPI processes such as $W^{\pm}j+W^{\pm}j$, $W^{\pm}j+Zj$, or $Zj+Zj$ are simulated with {\textsc pythia} 8 with CTEQ6L1 and the overall contribution is found to be negligible.

\subsection{Non-prompt background}
Non-prompt backgrounds come from processes with jets misidentified as leptons or leptons from hadron decays 
(including $b$- and $c$-hadron decays). Since the MC simulation may not accurately model the details of these processes, 
a data-driven fake-factor method is employed to estimate this contribution. 

The fake-factor method estimates a fake factor using the ratio of the number of jets satisfying the tight lepton identification criteria to the 
number of jets satisfying the loose lepton identification criteria in a jet-enriched sample. 
A new data sample, referred to as the ``tight$+$loose'' sample,  is selected with the same set of criteria as 
the signal region but one lepton is required to be a loose lepton. This sample is dominated by contributions from $W+$jets, $t\bar{t}$, and single-top-quark processes. 
The fake factor is measured, as discussed below, as a function of the loose lepton $\pt$ and applied to the tight+loose sample event-by-event as a global event weight  
to estimate the non-prompt background.
The contribution from multijet background with two jets satisfying the tight lepton requirements is estimated by selecting events with two loose leptons 
and using the product of the two factors computed for each lepton as the event weight. 
The contribution from multijet background is found to be less than 3.5\% of the total non-prompt background.

The lepton fake factors are measured using a dijet sample.
Events are selected with a `tag' jet and a loose or tight lepton back-to-back in the azimuthal plane with $\Delta \phi(\ell, j)>2.8$. The lepton is also referred to as an `underlying jet' since it originates from a jet or hadronic decay.
Both the lepton and the jet are required to have $p_{\textrm T}>$ 25 \GeV. 
The transverse mass of the lepton and $E_{\textrm T}^{\textrm miss}$ system is required to be less than 40~\GeV\ to suppress the $W+$jets contamination. 
The tag jet and underlying jet recoil in the transverse plane and are assumed to have the same \pt. The underlying jet \pt\ is calculated as the sum of the lepton \pt\ plus the transverse energy deposited in a cone of radius $\Delta 
R$ < 0.3 around the lepton. To account for the reduction in \pt\ from energy deposited outside the lepton isolation cone or loss due to neutrinos, 
the tag jet \pt\ distribution in the dijet sample is reweighted 
to match the underlying jet in the tight+loose sample. 
The energy loss is linearly dependent on \pt\ where the tag jet has 18\% higher \pt\ than the underlying jet associated with an electron and 72\% more for underlying jets associated with a muon. The energy loss for non-prompt muons is accountable by the loss from neutrinos given these events are derived mainly from $c$- and $b$-hadron decays.
In addition, a correction factor is applied to the tight+loose sample to take into account 
the lower trigger efficiency of isolated lepton triggers for loose leptons. The final fake factors are on the order of 2\% for electrons and less than 1\% for muons.

\subsection{Conversion background}
The conversion background is divided into two categories: events containing 
two prompt leptons with opposite electric charge, which can mimic the same 
final state if the electric charge of one lepton is misidentified (denoted by ``Charge misID''), and 
$W\gamma$ production with the photon misreconstructed as an electron (denoted by ``$W\gamma$''). 

The dominant mechanism responsible for charge misidentification of prompt electrons is the radiation of an energetic photon, which subsequently converts into an $e^+e^-$ pair. The charge misidentification rate for muons is negligible and is therefore not considered. Events entering the signal regions due to conversions consist mainly of fully leptonic $t\bar{t}$ decays and Drell--Yan 
lepton pair production.

The rate of electron charge misidentification is measured in a data sample enriched in $Z \rightarrow e^+ e^-$ events. 
This sample is required to have two tight electrons with the dielectron invariant mass between 70 \GeV\ 
and 100 \GeV. The asymmetric window around the pole mass of the $Z$ boson is used to account for the reduced reconstructed energy when an electron's charge is misidentified. Contributions to this mass region from other processes are found to be less than 1\%. 
No requirement is made on the charges of the two electrons. 
The per-electron misidentification rate is derived from the number of same-electric-charge events and the total number of dielectron events.

A likelihood fit is used to measure the charge misidentification 
rate as a function of the electron $\pt$ and $\eta$, taking into account that either electron in a same-electric-charge pair 
could be the misidentified one. The numbers of dielectron events and same-electric-charge 
events are counted in bins of the electron $\pt$ and $\eta$. While the process of charge misidentification is inherently binomial, given the large number of events and the relatively small charge-flip rate a Poisson distribution is assumed. Given the total number of observed dielectron events, $N^{i,j}$, and the charge misidentification rates, $\epsilon^i$ and $\epsilon^j$, where the efficiency is given for bins of \pT{} and $\eta$ for the two electrons, $i$ and $j$, the expected number of same-electric-charge events ($\tilde{N}_{SS}^{i,j}$) is given by

\begin{equation}
  \tilde{N}_\mathrm{SS}^{i,j} = \left[ \epsilon^i \left( 1 - \epsilon^j \right) + \epsilon^j \left( 1 - \epsilon^i \right) \right] N^{i,j} \approx \left(\epsilon^i + \epsilon^j \right) N^{i,j} \ .
\end{equation}

The approximation is valid for very small charge misidentification rates. The log-likelihood function for the number of observed 
dielectron events with same electric charge ($N_\mathrm{SS}^{i,j}$) with respect to an expectation of $  \tilde{N}_\mathrm{SS}^{i,j}$  is therefore given by

\begin{equation}
 \ln L_\mathrm{misID} = \ln \prod \limits_{i,j} \frac{\left[\left(\epsilon^i + \epsilon^j \right) N^{i,j}\right]^{N_\mathrm{SS}^{i,j}}}{N_\mathrm{SS}^{i,j}!} e^{-\left(\epsilon^i + \epsilon^j \right) N^{i,j}} = \sum\limits_{i,j} \left[ N_\mathrm{SS}^{i,j} \ln  N^{i,j}(\epsilon^{i} + \epsilon^{j}) - N^{i,j}(\epsilon^{i} + \epsilon^{j}) - \ln  {N_\mathrm{SS}^{i,j}!} \right].
  \label{eq:charge_llh}
\end{equation} 
Charge misidentification rates are determined for each $\pt$ and $\eta$ bin by 
maximizing the above log-likelihood function given the observed counts.
Since the rates for bremsstrahlung  and photon conversion depend on the amount of material traversed, the charge misidentification rate 
exhibits a strong dependence on the $\eta$ of the electron with the rate generally increasing with $|\eta|$. 
The charge misidentification rate is observed to be a few tenths of a percent over most of the $\eta$ range with a maximum of about 2$\%$ near $|\eta|=$2.5.

The measured electron charge misidentification rate is cross-checked using a tag-and-probe method applied to the $Z \rightarrow e^+ e^-$ sample. Tighter 
requirements on the quality of the cluster in the calorimeter and the matched track are imposed on the tag electron to make sure its electric charge is correctly determined. 
The electric charge of the second electron is used to measure the electron charge misidentification rate. Good agreement between the estimates 
from these two methods is found.   

To predict the amount of background from charge misidentification, data events are selected 
using all of the signal region criteria but requiring the two leptons to have opposite-sign electric charges. 
For each electron in this data sample, the corresponding charge misidentification rate is included in the global event weight. 
In the case of events with two electrons, this procedure is applied to each electron separately. 
In addition, an energy correction is applied to the electron with the charge misidentification rate assigned to take into account that electrons with misidentified charge tend to have lower reconstructed energy than their correctly identified counterparts and also yield a wider dielectron invariant mass peak for the $Z$ boson.
This energy correction is determined using the electron generator-level and reconstructed energies in MC-simulated $Z \rightarrow e^+ e^-$ events.

Production of $W\gamma$ events can yield same-electric-charge leptons if the photon 
converts in the detector and one conversion electron is not reconstructed. 
Both electroweak and strong $W\gamma jj$ production can arise and their 
contributions are also estimated using MC-simulated samples. The electroweak production is estimated using 
{\textsc sherpa}, while the strong production is estimated using {\textsc alpgen}~\cite{alpgen}. The CTEQ6L1 PDF set is used for both samples.

\subsection{Control regions}
Four control regions (CRs), referred to as the ``$\leq$ 1 jet CR'', ``trilepton CR'', ``$b$-tag CR'', and ``low-$m_{jj}$ CR'', are 
used to validate background predictions. 
For all CRs, the contributions from \sswwjj-EW and \sswwjj-QCD production are normalized to the SM prediction.
The definitions of all four control regions, the number of observed data events and the 
SM predictions as well as a few kinematic distributions  in each region are presented below. 
The comparison between the data and the prediction is checked using a $\chi^2/$ndf test and good agreement is observed.

\begin{figure*}[pbht!]
\centering
\begin{tabular}{cc}
\centering
\includegraphics[width=0.48\textwidth]{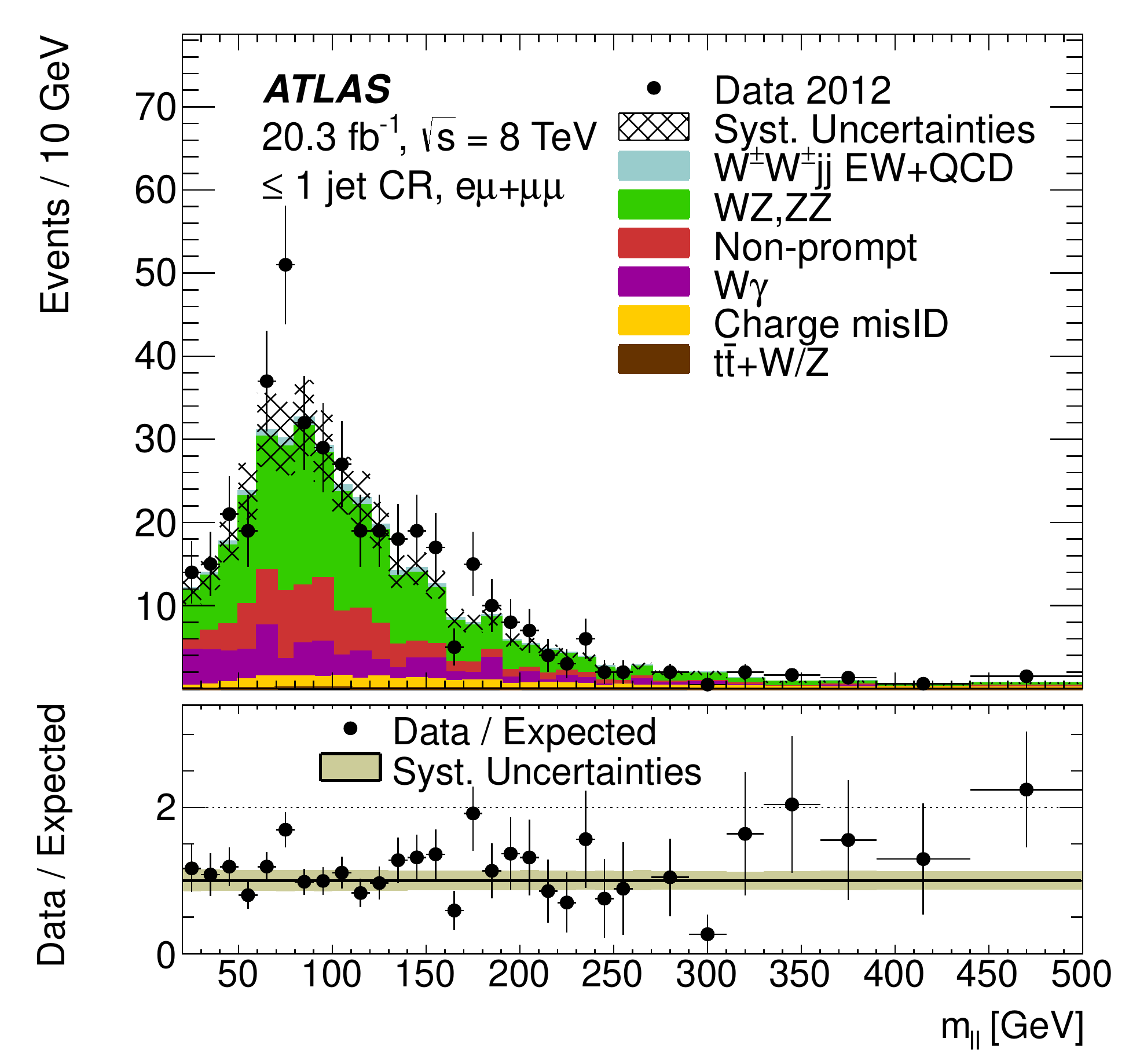} 
\includegraphics[width=0.48\textwidth]{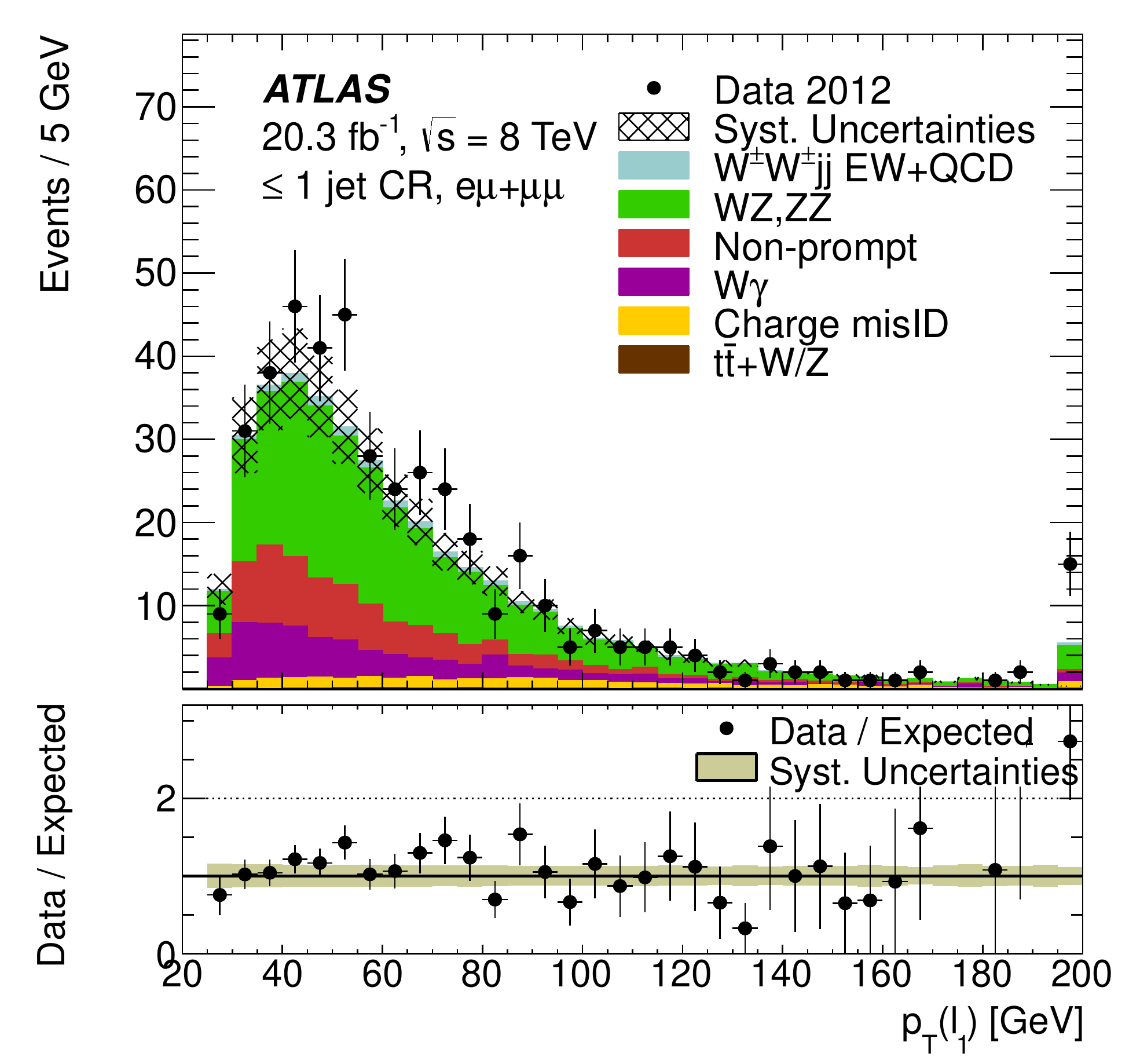} 
\end{tabular}
\caption{The invariant mass distribution of the dilepton pair (left) and the leading-lepton \pT\  distribution (right) for the $e^\pm \mu^\pm$ and 
$\mu^\pm \mu^\pm$ channels in the $\leq 1$ jet CR without the $Z$ boson veto requirement. 
 The error bars on the data points include statistical uncertainty only. The hatched band represents the systematic uncertainty of the total prediction. The lower plot shows the ratio of the data to the expected background where the brown band indicates the systematic uncertainty including the MC statistical uncertainty.
The last bin includes overflow events.} 
\label{fig:lowNjetEMMM}
\end{figure*}

\noindent\textbf{$\leq$ 1 jet control region}\newline
The $\leq$ 1 jet CR is used to test 
the modeling of lepton kinematics in the $WZ/ZZ$ background where one of the leptons from the $Z$ boson decay is 
not reconstructed. It is defined by inverting the signal region selection on the jet multiplicity to accept only events 
with at most one jet. As a consequence, selection criteria using jet-based quantities such as $m_{jj}$ and $\Delta y_{jj}$ are also dropped. 
Figure~\ref{fig:lowNjetEMMM} shows the dilepton invariant mass distribution and the leading-lepton \pT\ distribution for the $e^\pm \mu^\pm$ and $\mu^\pm \mu^\pm$ channels with the $Z$ boson veto dropped.  
Table~\ref{tab:lowNjetCR} shows the number of data events compared to the predictions from signal and various background sources. 

\begin{table}[h]
\centering
\begin{tabular}{ll|ccc|c}
\hline
\hline
\multicolumn{6}{c}{$\leq1$ jet Control Region}\\
\hline
&&$e^\pm e^\pm$&$e^\pm \mu^\pm$&$\mu^\pm \mu^\pm$&Total\\
\hline
 \multicolumn{2}{l|}{\sswwjj-EW+QCD} &\ \ 2.2 $\pm$ 0.3  &\ \ 7.0 $\pm$ 0.7&4.5 $\pm$ 0.5&13.7 $\pm$ 1.4\\ \hline
 \multirow{2}{*}{Prompt}  & $WZ$,$ZZ$ &46 $\pm$ 8&130 $\pm$ 23&75 $\pm$ 13&250 $\pm$ 40\\ \cline{2-6}
 & $t\bar{t}$+$W$/$Z$&\ \ 0.3 $\pm$ 0.2&\ \ 0.8 $\pm$ 0.4&0.6 $\pm$ 0.3&\ \ 1.7 $\pm$ 0.7\\ \hline
 \multirow{2}{*}{Conversions} & Charge misID &152 $\pm$ 17&24 $\pm$ 4&--&177 $\pm$ 21\\ \cline{2-6}
 & $W\gamma$&\ \ 39 $\pm$ 11 &\ \ 59 $\pm$ 17&0.04 $\pm$ 0.04&\ \ 98 $\pm$ 29\\ \hline 
 \multicolumn{2}{l|}{Non-prompt} &\ \ 38 $\pm$ 15&\ \ 65 $\pm$ 26&8 $\pm$ 5&111 $\pm$ 30\\  \hline
 \multicolumn{2}{l|}{Total predicted} &278 $\pm$ 28&290 $\pm$ 40&88 $\pm$ 14&650 $\pm$ 70\\  \hline
 \multicolumn{2}{l|}{Data} &288&328&101&717\\
\hline
\hline
\end{tabular}
\caption{Predicted and observed numbers of events in the $\leq1$ jet control region separately for the \elel, \elmu, and \mum channels as well as for the sum of all three. The uncertainty is the combination of statistical and systematic uncertainties; correlations among systematic uncertainties are taken into account in the calculation of the total.
}
\label{tab:lowNjetCR}
\end{table} 

\begin{figure*}[pbht!]
\centering
\begin{tabular}{cc}
\includegraphics[width=0.48\textwidth]{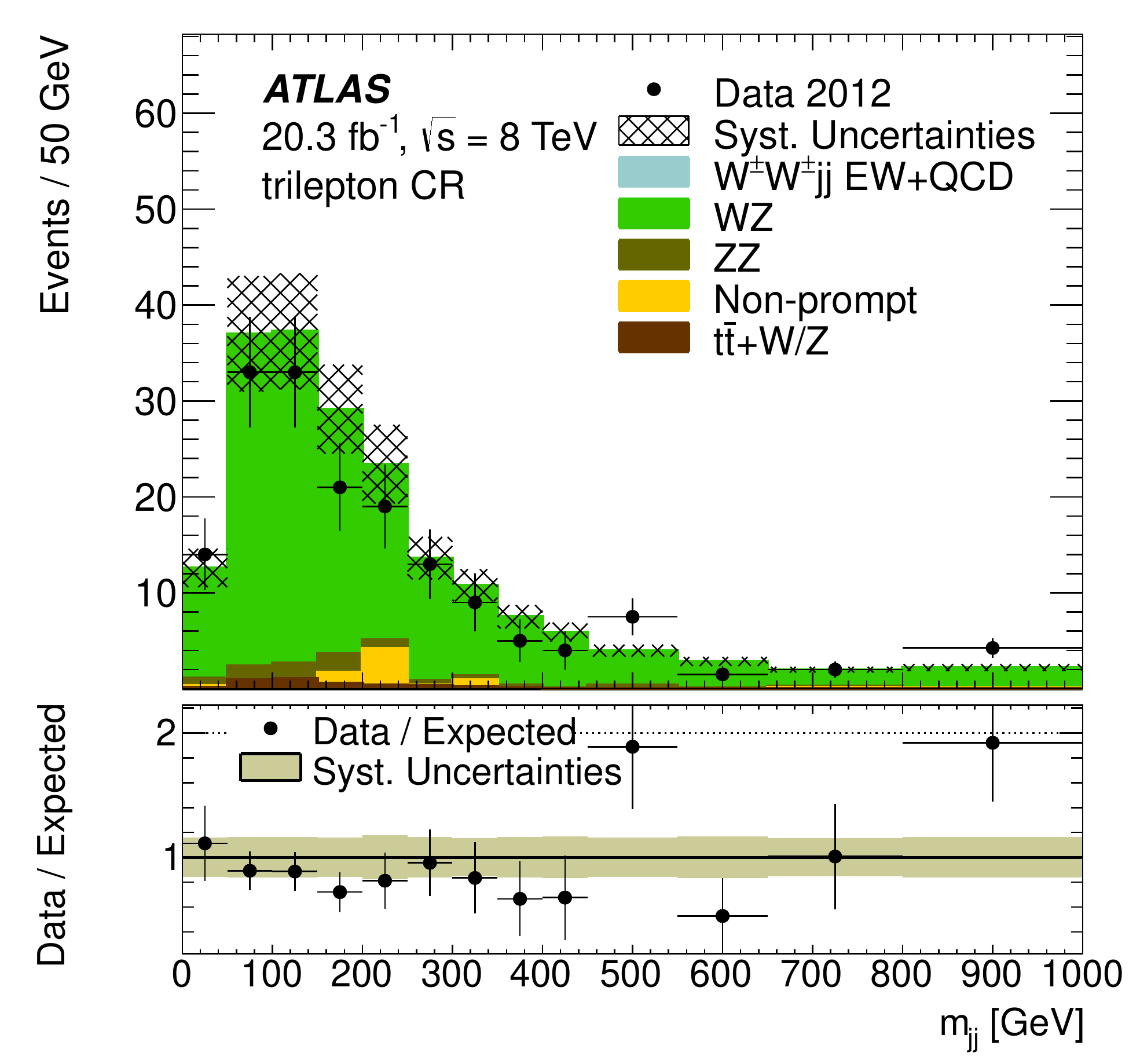} 
\includegraphics[width=0.48\textwidth]{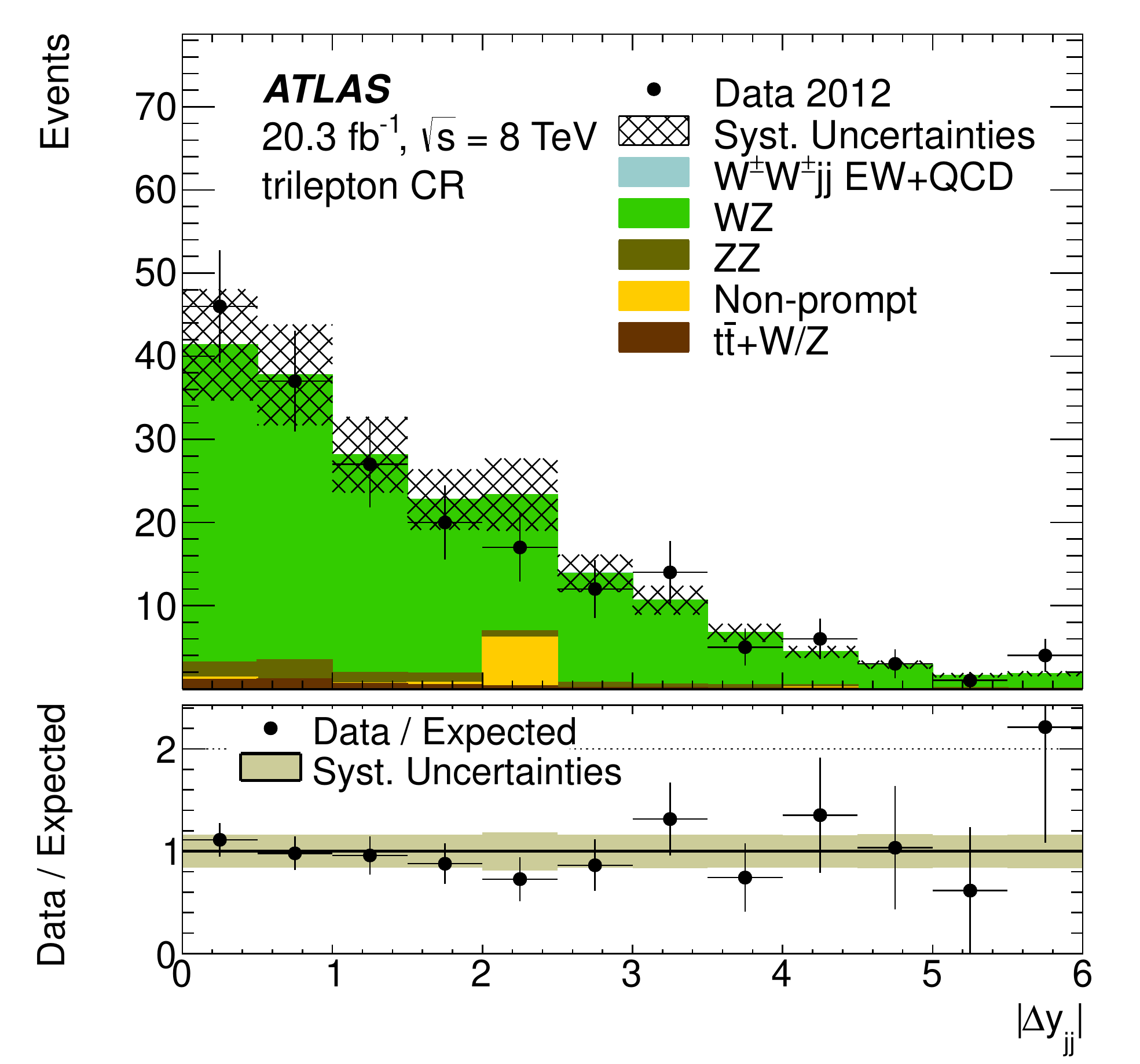}
\end{tabular}
\caption{The \mjj distribution (left) and the distribution of the difference in rapidity (right) of the two jets with the highest $\pt$ is shown summed over all lepton channels for the trilepton CR. 
Non-prompt background in this region is estimated using MC simulation. 
 The error bars on the data points include statistical uncertainty only. The hatched band represents the systematic uncertainty of the total prediction. The lower plot shows the ratio of the data to the expected background where the brown band indicates the systematic uncertainty including the MC statistical uncertainty.
The last bin includes overflow events.}
\label{fig:trilep}
\end{figure*}

\noindent\textbf{Trilepton control region} \\
The trilepton CR provides a test of the modeling of lepton and jet kinematics of the $WZjj$ production.
It is defined by selecting events with three charged leptons where the third lepton passes the veto-lepton requirements. 
Events containing a fourth lepton passing the veto-lepton definition are still rejected. 
In contrast, $m_{jj}$ and $\Delta y_{jj}$ selection criteria are also dropped to obtain more events. 
The $m_{jj}$ and $|\Delta y_{jj}|$ distributions are shown in Figure~\ref{fig:trilep}. 
Table~\ref{tab:TrileptonCRtab} shows the number of data events compared to the predictions from signal and various background sources. 
\begin{table}[h!]
\centering
\begin{tabular}{ll|ccc|c}
\hline
\hline
\multicolumn{6}{c}{Trilepton Control Region}\\
\hline
& &$e^{\pm}e^{\pm}\ell^{\mp}$&$e^{\pm}\mu^{\pm}\ell^{\mp}$&$\mu^{\pm}\mu^{\pm}\ell^{\mp}$&Total\\
\hline
 \multicolumn{2}{l|}{\sswwjj-EW+QCD} &\ \ 0.05 $\pm$ 0.02&\ \ 0.13 $\pm$ 0.03& --&\ \ 0.168 $\pm$ 0.029\\  \hline
\multirow{3}{*}{Prompt} & $WZ$ &32 $\pm$ 5&\ \ 96 $\pm$ 16&57 $\pm$ 10&186 $\pm$ 31\\ \cline{2-6}
&  $ZZ$ &\ \ 2.2 $\pm$ 0.6&\ \ 5.3 $\pm$ 1.3&1.8 $\pm$ 0.5&\ \ 9.2 $\pm$ 2.1\\ \cline{2-6}
&  $t\bar{t}$+$W$/$Z$&\ \ 0.7 $\pm$ 0.3&\ \ 2.4 $\pm$ 1.0&1.0 $\pm$ 0.5&\ \ 4.1 $\pm$ 1.7\\ \hline
\multicolumn{2}{l|}{Non-prompt} &\ \ 0.5 $\pm$ 0.3&\ \ 4 $\pm$ 4& -- &\ \ 4 $\pm$ 4\\  \hline
 \multicolumn{2}{l|}{Total predicted}  &36 $\pm$ 6&108 $\pm$ 18&60 $\pm$ 10&204 $\pm$ 33\\ \hline
 \multicolumn{2}{l|}{Data} &40&104&48&192\\
\hline
\hline
\end{tabular}

\caption{Predicted and observed numbers of events in the trilepton control region separately for the \elel, \elmu, and \mum channels as well as for the sum of all three. 
The third lepton is required to pass the veto-lepton requirements. 
The uncertainty is the combination of statistical and systematic uncertainties; correlations among systematic uncertainties are taken into account in the calculation of the total. 
The conversion background is found to be negligible.}
\label{tab:TrileptonCRtab}
\end{table}

\begin{figure*}[ht!]
\centering
\begin{tabular}{cc}
\includegraphics[width=0.48\textwidth]{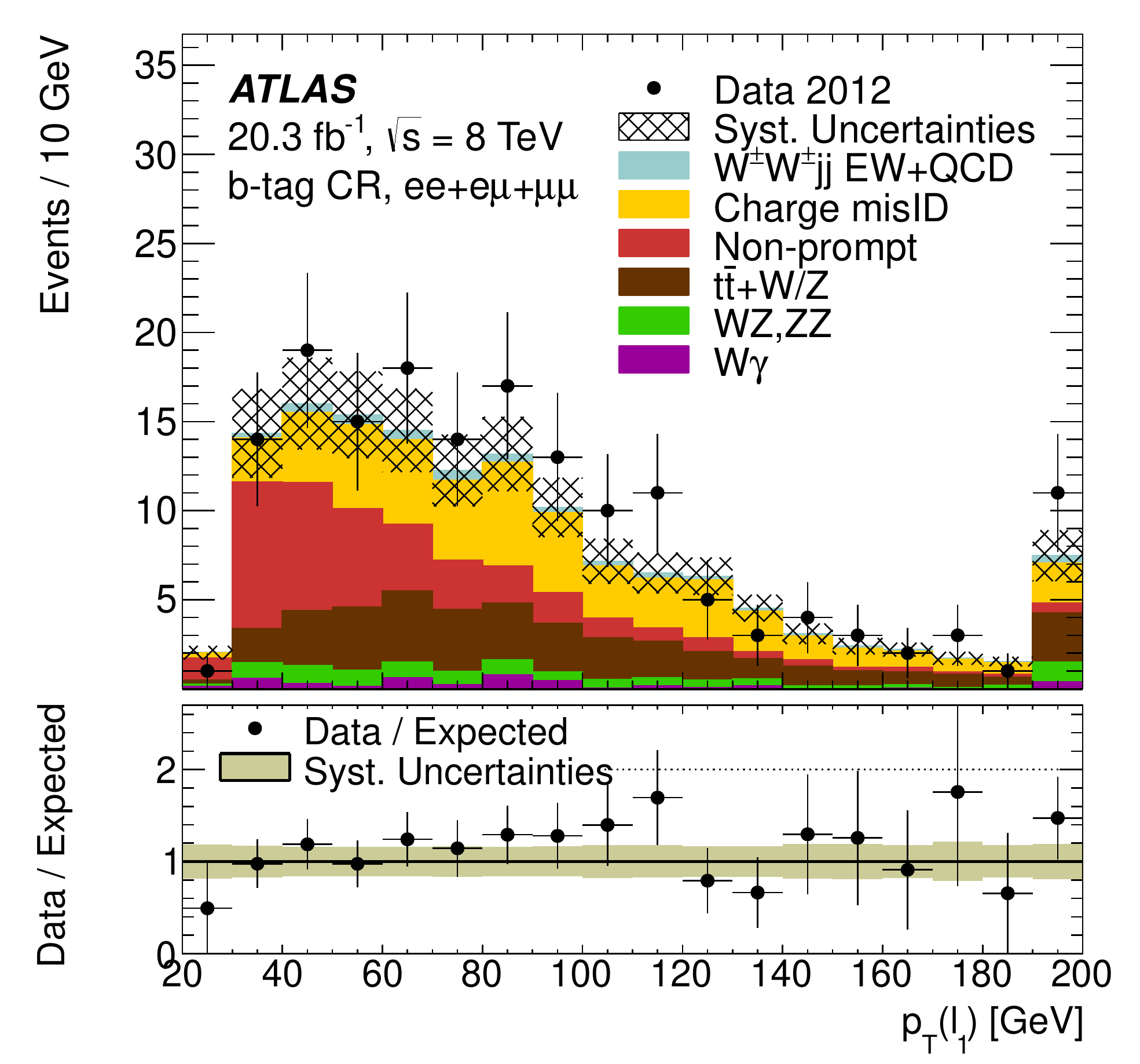}
\includegraphics[width=0.48\textwidth]{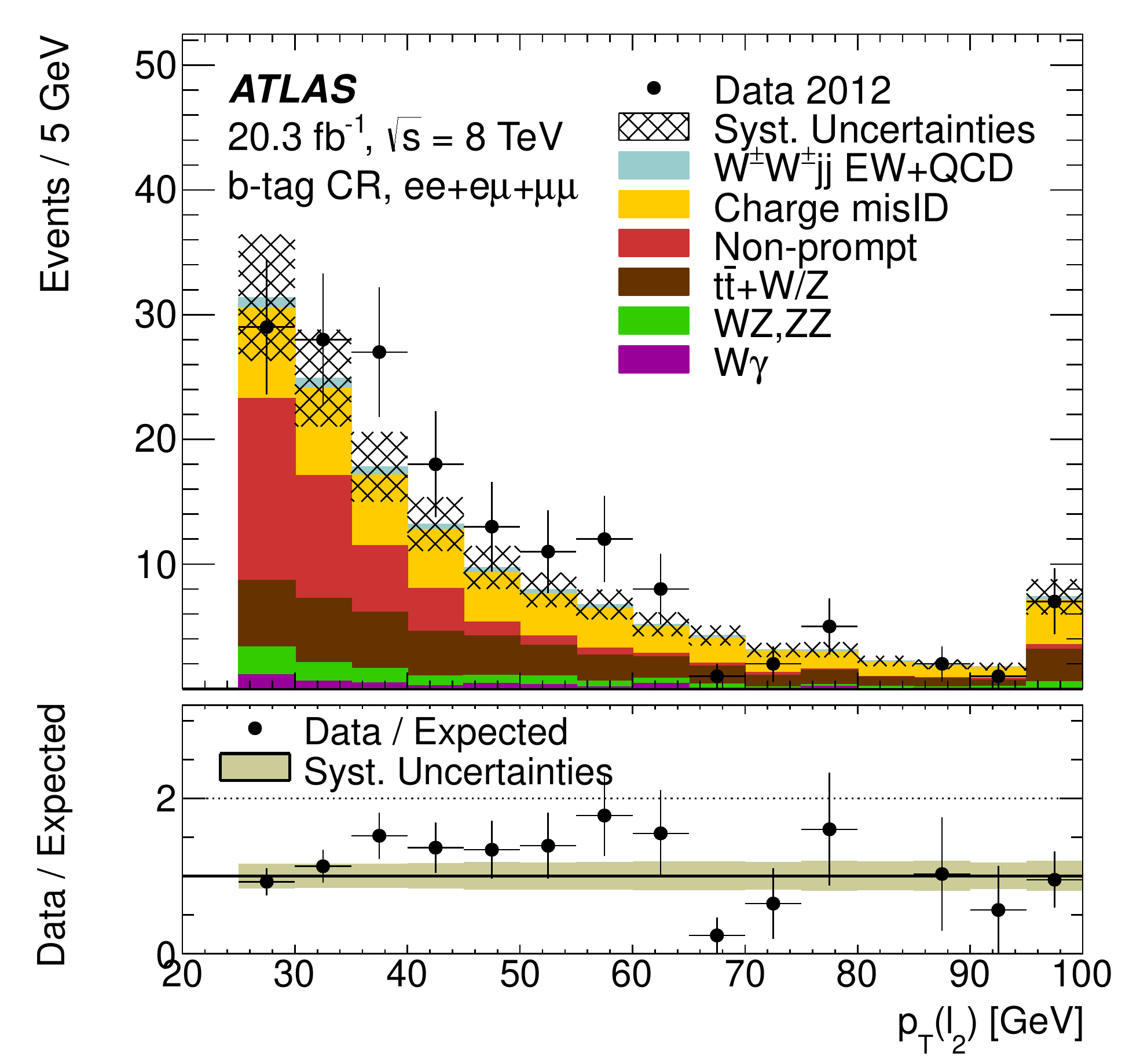}
\end{tabular}
\caption{The leading (left) and sub-leading (right) lepton $\pt$ distribution in the $b$-tag CR. 
The conversions background has been split into $W\gamma$ events and events with two prompt, opposite-sign~(OS) leptons. 
 The error bars on the data points include statistical uncertainty only. The hatched band represents the systematic uncertainty of the total prediction. The lower plot shows the ratio of the data to the expected background where the brown band indicates the systematic uncertainty including the MC statistical uncertainty.
The last bin includes overflow events.}
\label{fig:btagLepPt} 
\end{figure*}

\noindent\textbf{$b$-tag control region} \\
The $b$-tag CR provides a test of the modeling of $t\bar{t}+W/Z$ and non-prompt background.
It is defined by inverting the $b$-jet veto criteria to require the presence of at least one $b$-tagged jet in the event. 
The $m_{jj}$ and $|\Delta y_{jj}|$ selection criteria are also dropped. 
Transverse momentum distributions for the leading- and sub-leading-leptons are shown in Figure~\ref{fig:btagLepPt}.
Table~\ref{tab:btagCRtab} shows the number of data events compared to the predictions from signal and various background sources. The $b$-tagging efficiency is included in the systematic uncertainty described in Section~\ref{sec:systematics}. 
\begin{table}[h!]
\centering
\begin{tabular}{ll|ccc|c}
\hline
\hline
\multicolumn{6}{c}{$b$-tag Control Region}\\
\hline
& &$e^\pm e^\pm$&$e^\pm \mu^\pm$&$\mu^\pm \mu^\pm$&Total\\
\hline
 \multicolumn{2}{l|}{\sswwjj-EW+QCD} &\ \ 0.8 $\pm$ 0.1&\ \ 2.6 $\pm$ 0.3&\ \ 1.5 $\pm$ 0.2&\ \ 4.9 $\pm$ 0.5\\ \hline
\multirow{2}{*}{Prompt} & $WZ$,$ZZ$ &\ \ 2.3 $\pm$ 0.5&\ \ 4.9 $\pm$ 0.9&\ \ 2.2 $\pm$ 0.4&\ \ 9.4 $\pm$ 1.6\\ \cline{2-6}
& $t\bar{t}$+$W$/$Z$&\ \ 7.1 $\pm$ 3.1&18 $\pm$ 8&11 $\pm$ 4&\ \ 36 $\pm$ 15\\ \hline
\multirow{2}{*}{Conversions} & Charge misID &22 $\pm$ 5&27 $\pm$ 6&--&\ \ 49 $\pm$ 11\\  \cline{2-6}
& $W\gamma$&\ \ 1.7 $\pm$ 0.7&\ \ 2.3 $\pm$ 0.9&\ \ 0.2 $\pm$ 0.2&\ \ 4.2 $\pm$ 1.4\\ \hline
 \multicolumn{2}{l|}{Non-prompt} &\ \ 6.7 $\pm$ 2.5&20 $\pm$ 8&10 $\pm$ 5&\ \ 37 $\pm$ 10\\ \hline
 \multicolumn{2}{l|}{Total predicted} &41 $\pm$ 6&\ \ 75 $\pm$ 13&25 $\pm$ 7&141 $\pm$ 22\\ \hline
 \multicolumn{2}{l|}{Data} &46&82&36&164\\
\hline
\hline
\end{tabular}
\caption{Predicted and observed numbers of events in the $b$-tag control region separately for the \elel, \elmu, and \mum channels as well as for the sum of all three. The uncertainty is the combination of statistical and systematic uncertainties; correlations among systematic uncertainties are taken into account in the calculation of the total. } 
\label{tab:btagCRtab}
\end{table}

\begin{figure*}[ht!]
\centering
\includegraphics[width=0.48\textwidth]{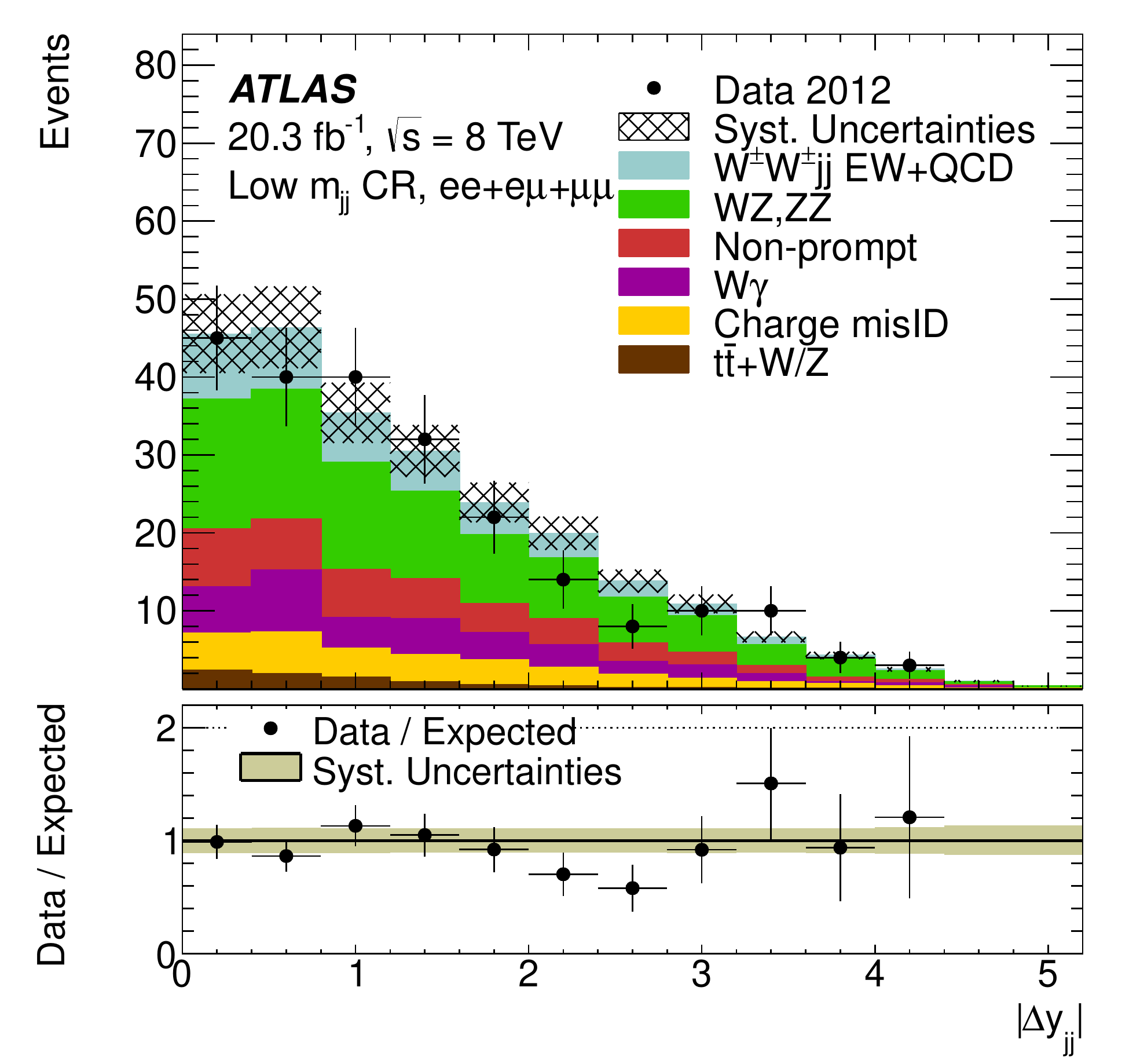} 
\includegraphics[width=0.48\textwidth]{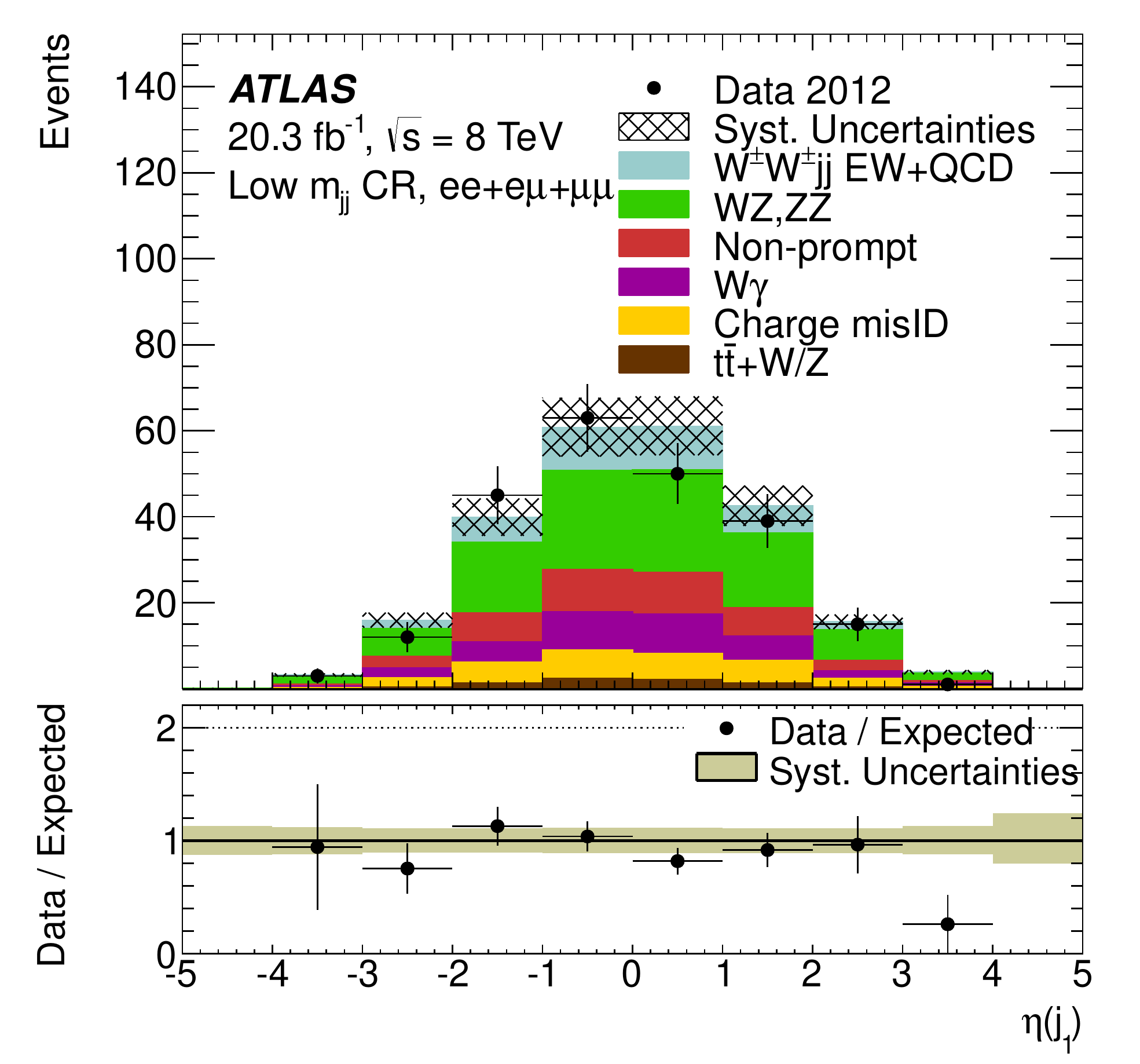}
\caption{The distribution of the rapidity difference between the two jets with the highest $\pt$ (left) and the distribution of the $\eta$ of the leading-jet (right) for the sum of events in the \elel, \elmu, and \mum channels for the low-\mjj\ CR. 
The conversions background has been split into $W\gamma$ events and events with two prompt OS leptons. 
 The error bars on the data points include statistical uncertainty only. The hatched band represents the systematic uncertainty of the total prediction. The lower plot shows the ratio of the data to the expected background where the brown band indicates the systematic uncertainty including the MC statistical uncertainty.
The last bin includes overflow events.}
\label{fig:lowMjjCR} 
\end{figure*}

\noindent\textbf{Low-$m_{jj}$ control region} \\
The low-$m_{jj}$ control region is used to check the background modeling in a
region with background composition similar to the signal regions. 
It is defined by inverting the $m_{jj}$ selection and dropping the $|\Delta y_{jj}|$ selection. 
The $|\Delta y_{jj}|$ and leading-jet $\pt$ distributions in the low-$m_{jj}$ control region are shown in Figure~\ref{fig:lowMjjCR}. 
Table~\ref{tab:lowMjjCRtab} shows the number of data events compared to the predictions from signal and various background sources. 
\begin{table}
\centering
\begin{tabular}{ll|ccc|c}
\hline
\hline
\multicolumn{6}{c}{Low $m_{jj}$ Control Region}\\
\hline
& &$e^\pm e^\pm$&$e^\pm \mu^\pm$&$\mu^\pm \mu^\pm$&Total\\
\hline
 \multicolumn{2}{l|}{\sswwjj-EW+QCD} &\ \ 5.9 $\pm$ 0.6&17.4 $\pm$ 1.8&~~10.6 $\pm$ 1.1&33.9 $\pm$ 3.4\\ \hline
 \multirow{2}{*}{Prompt} & $WZ$,$Z$Z&25 $\pm$ 4&54 $\pm$ 9&~~18.4 $\pm$ 3.1&\ \ 98 $\pm$ 16\\ \cline{2-6}
 & $t\bar{t}$+$W$/$Z$&\ \ 1.7 $\pm$ 0.7&\ \ 3.8 $\pm$ 1.6&\ \ ~~2.4 $\pm$ 1.0&\ \ 7.9 $\pm$ 3.4\\ \hline
 \multirow{2}{*}{Conversions}  & Charge misID &19.4 $\pm$ 2.3&\ \ 8.4 $\pm$ 1.4&--&27.8 $\pm$ 3.4\\ \cline{2-6}
 & $W\gamma$&14 $\pm$ 4&20 $\pm$ 6&--&\ \ 34 $\pm$ 10\\ \hline
 \multicolumn{2}{l|}{Non-prompt} &\ \ 9 $\pm$ 4&21 $\pm$ 8&~~8 $\pm$ 4&\ \ 39 $\pm$ 10\\ \hline
 \multicolumn{2}{l|}{Total predicted} &75 $\pm$ 9&125 $\pm$ 16&39 $\pm$ 6&240 $\pm$ 27\\ \hline
 \multicolumn{2}{l|}{Data} &78&120&30&228\\
\hline
\hline
\end{tabular}
\caption{Predicted and observed numbers of events in the low-$m_{jj}$ control region separately for the \elel, \elmu, and \mum channels as well as for the sum of all three. The uncertainty is the combination of statistical and systematic uncertainties; correlations among systematic uncertainties are taken into account in the calculation of the total. } 
\label{tab:lowMjjCRtab}
\end{table}

\section{Systematic uncertainties}
\label{sec:systematics}
Systematic uncertainties in the measured cross sections arise from uncertainties in the physics object reconstruction and identification, 
the procedures used to correct for detector effects, the background estimation, the usage of theoretical cross-sections for signal and background processes, and luminosity. 

The experimental systematic uncertainties affecting the signal and prompt-background estimates include: the uncertainties due to the lepton energy 
scale, energy resolution, and identification efficiency~\cite{elecID, muonID}; the uncertainties due to the jet energy scale and resolution, which include the pileup jet uncertainty contribution at roughly 25\% of the total jet systematic uncertainty~\cite{jetEnergy}; 
the uncertainties in the \met\ calculation from energy deposits not associated with reconstructed objects~\cite{met}; and the uncertainties due to $b$-tagging 
efficiency and mistag rate~\cite{bjet}. 
An uncertainty is applied to MC samples to cover differences in efficiency observed between the trigger in data and the MC trigger simulation.
The uncertainty in the integrated luminosity is 2.8\%, affecting the overall normalization of both the signal and background processes 
estimated from MC simulation. It is derived following the methodology detailed in Ref.~\cite{lumi}.

The uncertainty in the non-prompt-background estimate is between 39\% and 52\% depending on region and channel.
It is dominated by the prompt-lepton contamination in the dijet sample used to estimate the fake factors, the uncertainty in the extrapolation of fake factors into the signal region, and the statistical uncertainty in the number of ``tight+loose'' events used to estimate the background.

The dominant systematic uncertainties from the conversion background arise from a possible method bias and 
the statistical uncertainty in the charge misidentification rate measurement. The total uncertainty in the estimation of the 
conversion background is found to be between 15\% and 32\% depending on signal region and lepton flavor.

The dominant theoretical uncertainty in the prompt background estimation comes from the predicted cross-section uncertainties for the \wzjj-EW and \wzjj-QCD production. 
Systematic uncertainties in the \wzjj-EW background estimation are determined separately for the contribution with and without $b$-quarks.
Uncertainties due to the choice of factorization and renormalization scales and PDF uncertainties are calculated with {\textsc vbfnlo}. 
Parton-shower effects are determined by applying two parton showering algorithms. LO {\textsc vbfnlo} events are used, since no NLO events are available. 
The difference between the {\textsc pythia 8} parton-shower model with the AU2 tune for the underlying-event modeling
and the \herwig{} parton shower with {\textsc jimmy} for the underlying-event modeling is used to estimate the parton-shower uncertainty. 
The same procedures are used to calculate the total NLO cross-sections, scale, PDF, and parton-shower uncertainties for the \wzjj-QCD production. 
The \wzjj-QCD final state also occurs through diagrams with zero or one parton 
but containing two jets after parton showering. This contribution is included in the \sherpa\ sample and 
has an additional parton-shower uncertainty. This effect is determined using a dedicated \madgraph\ sample with two different 
parton-shower models. A 52\,\% uncertainty is obtained from this comparison, which results in an uncertainty of 6\,\% in the total \wzjj-QCD contribution. 
The theoretical uncertainties of the other background contributions include 30\%, 19\%, and 17\% uncertainties in the predicted cross-sections of the $t\bar{t}+V$, electroweak and strong production of $ZZjj$, and $W\gamma$ processes, respectively.

\begin{table*}[ht!]
\resizebox{\textwidth}{!}{ 
\begin{tabular}{lcc|cc}
\hline
\hline
\multicolumn{5}{c}{\centering Relative Systematic Uncertainties \ssee/\ssem/\ssmm{} [$\%$]}\\
\hline
~ & \multicolumn{2}{c|}{Background Yield}             & \multicolumn{2}{c}{Signal Yield}\\ 
\hline
& \phantom{0}Inclusive SR\phantom{0} & VBS SR             & \phantom{0}Inclusive SR\phantom{0} & VBS SR\\
\hline
\sswwjj-EW cross-section       & &       & 5 & 6 \\
\sswwjj-QCD cross-section & & \phantom{0} & 3.1 & -- \\
\wzjj-EW cross-section           & 6/8/11  & 5/5/8              &  & \\
\wzjj-QCD cross-section             & --          & 0.9/1.5/2.6 & & \\
MC statistics                               & 8/6/8 & 9/6/8 &  4/2.1/2.8 & 5/2.7/4\\  
Luminosity                                  & 1.7/2.1/2.4 & 1.7/2.1/2.4   & 2.8 & 2.8\\
Trigger efficiency                          & 0.1/0.2/0.4 & 0.1/0.2/0.4    & 0.1/0.3/0.5 & 0.1/0.3/0.5\\
Lepton reconstruction and identification                      & 1.6/1.2/1.2 & 1.7/1.1/1.1 & 1.9/1.0/0.7\phantom{0} & 1.9/1.0/0.7\\
Jet-related uncertainties                           & 11/13/13    & 13/20/20    & 6 & 5\\
\met\ reconstruction                        & 2.2/2.4/1.8 & 2.9/3.2/1.4            & 1.1 & 1.1\\
$b$-tagging efficiency                        & 1.0/1.1/1.0 & 0.8/0.9/0.7    & 0.6 & 0.6\\
Non-prompt               & 4/7/7 & 4/7/7 & &  \\
Conversions    & 6/4/--  & 6/4/--  & &  \\
$W\gamma$ cross-section               & 2.8/2.6/--  & 3.1/2.6/--  & &  \\
\hline
Total &  17/19/21 & 18/20/21  &  10/9/9 & 10/9/9 \\
\hline
\hline
\end{tabular}
}
\caption{ The decomposition of the relative systematic uncertainties in the estimated number of background and signal events for the Inclusive and VBS SRs. 
The left columns represent the uncertainties of the total background predictions in each channel from the listed source, while the right columns represent the uncertainties of the total signal predictions from each source. 
Three numbers in the same cell indicate the uncertainties for the $e^\pm e^\pm$, $e^\pm \mu^\pm$ and $\mu^\pm \mu^\pm$ channels, respectively. If only one 
number is present in a given cell, it means all three channels have the same systematic uncertainty.
}          
\label{tab:systIncSR}
\end{table*}

A summary of the decomposition of the systematic uncertainties in the estimated
number of background and signal events for the two SRs is given in
Table~\ref{tab:systIncSR}. Most uncertainties do not have an inherent dependence
on the flavor of the two leptons, but the size of the contribution to the total
background uncertainty does depend on the channel due to differences in the
composition of the background between channels. The fractional uncertainties
listed are quoted as the effect on the background yield or signal yield in
the \ssee, \ssem, and \ssmm\ channels separately. 
The largest uncertainty is the jet-related uncertainty for both the
signal and background estimations.

\section{Events yields in the signal regions}
\begin{figure*}[hpt!]
\centering
\includegraphics[width=.48\textwidth]{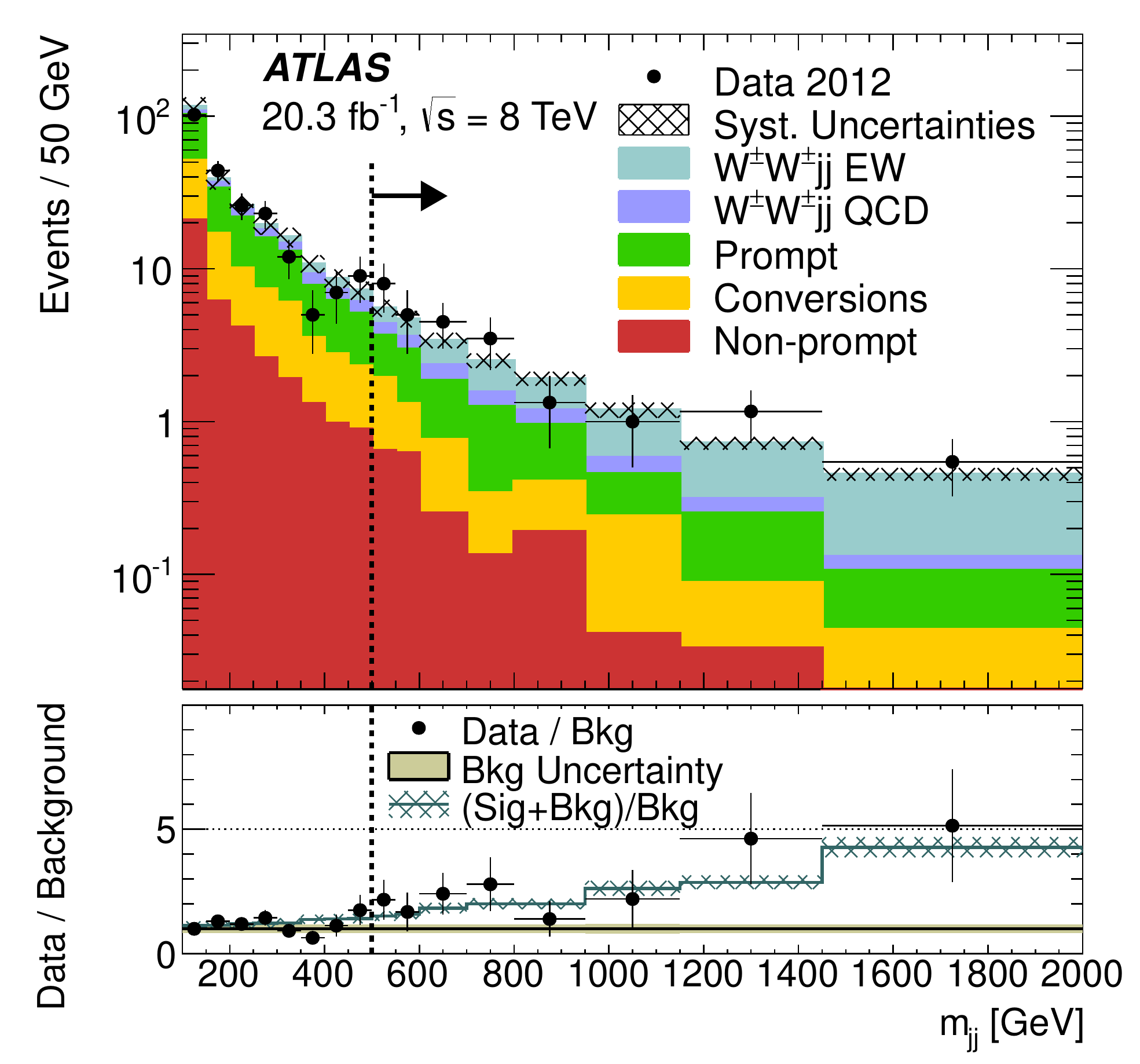}
\caption{The \mjj distribution for the combined channels in the Inclusive SR prior to applying the requirement that $m_{jj}>500$ \GeV. 
The error bars on the data points represent statistical uncertainty only. The hatched band represents the systematic uncertainty of the total prediction. 
The lower plot shows the ratio of the data to the expected background where the brown band indicates systematic uncertainty including the MC statistical uncertainty.
The ratio of the sum of the expected signal (\sswwjj-EW and \sswwjj-CQD) and background to the expected background is also shown.
}
\label{fig:SignalVBS:total:mjj}
\end{figure*}

\begin{figure*}[hpt!]
\centering
\includegraphics[width=.48\textwidth]{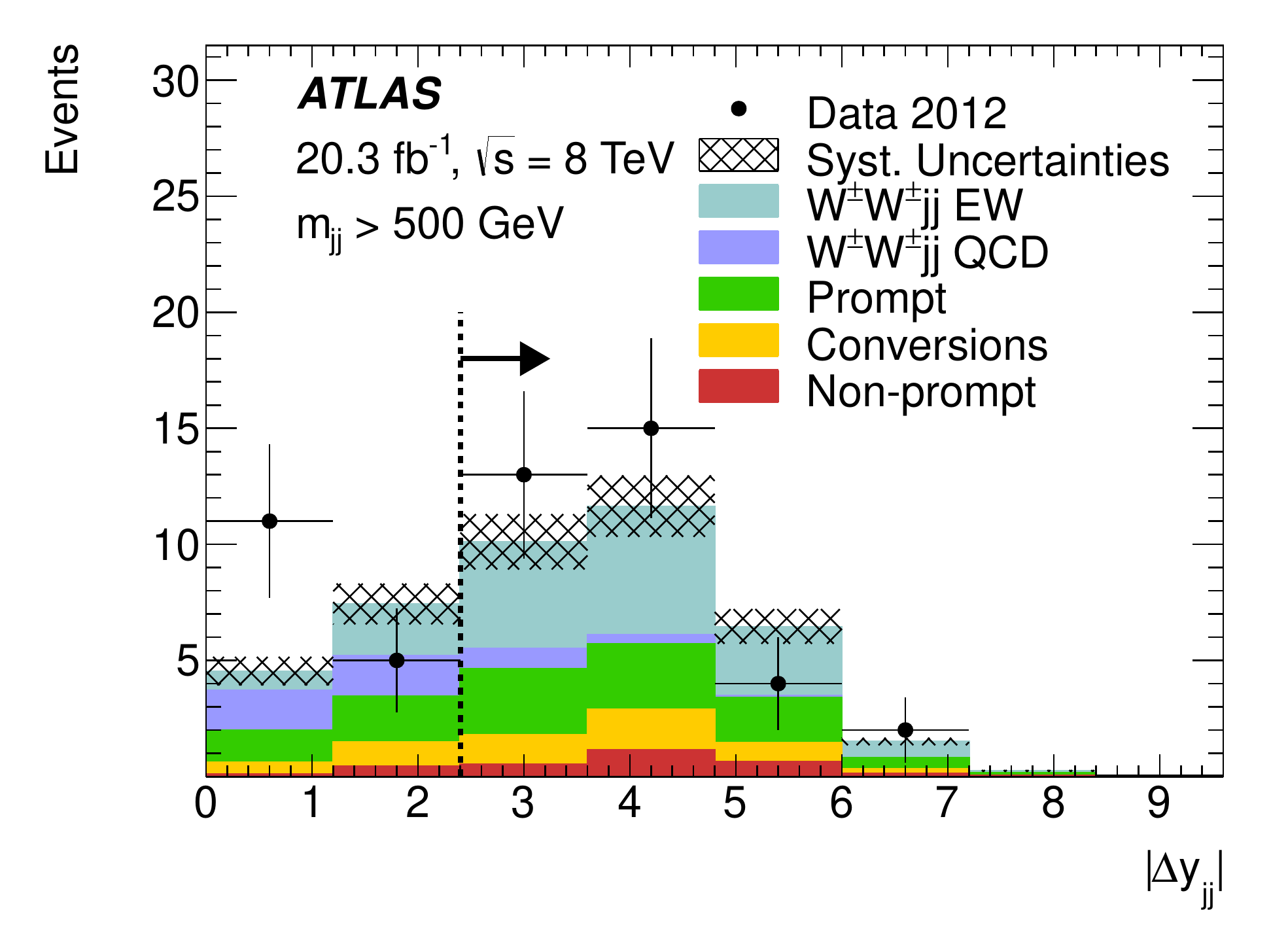}
\caption{The rapidity difference distribution between the two jets with the highest $\pt$ in the Inclusive SR for the combined channels. The region with \dyjj~$>$~ 2.4 denoted by the vertical dotted line indicates the VBS SR. The error bars on the data points include statistical uncertainty only. The hatched band represents the systematic uncertainty of the total prediction. 
The contributions from \sswwjj-EW and \sswwjj-QCD production are normalized to the SM prediction.} 
\label{fig:SignalVBS:total:dyjj}
\end{figure*}

\begin{figure*}[pht!]
\centering
\includegraphics[width=.48\textwidth]{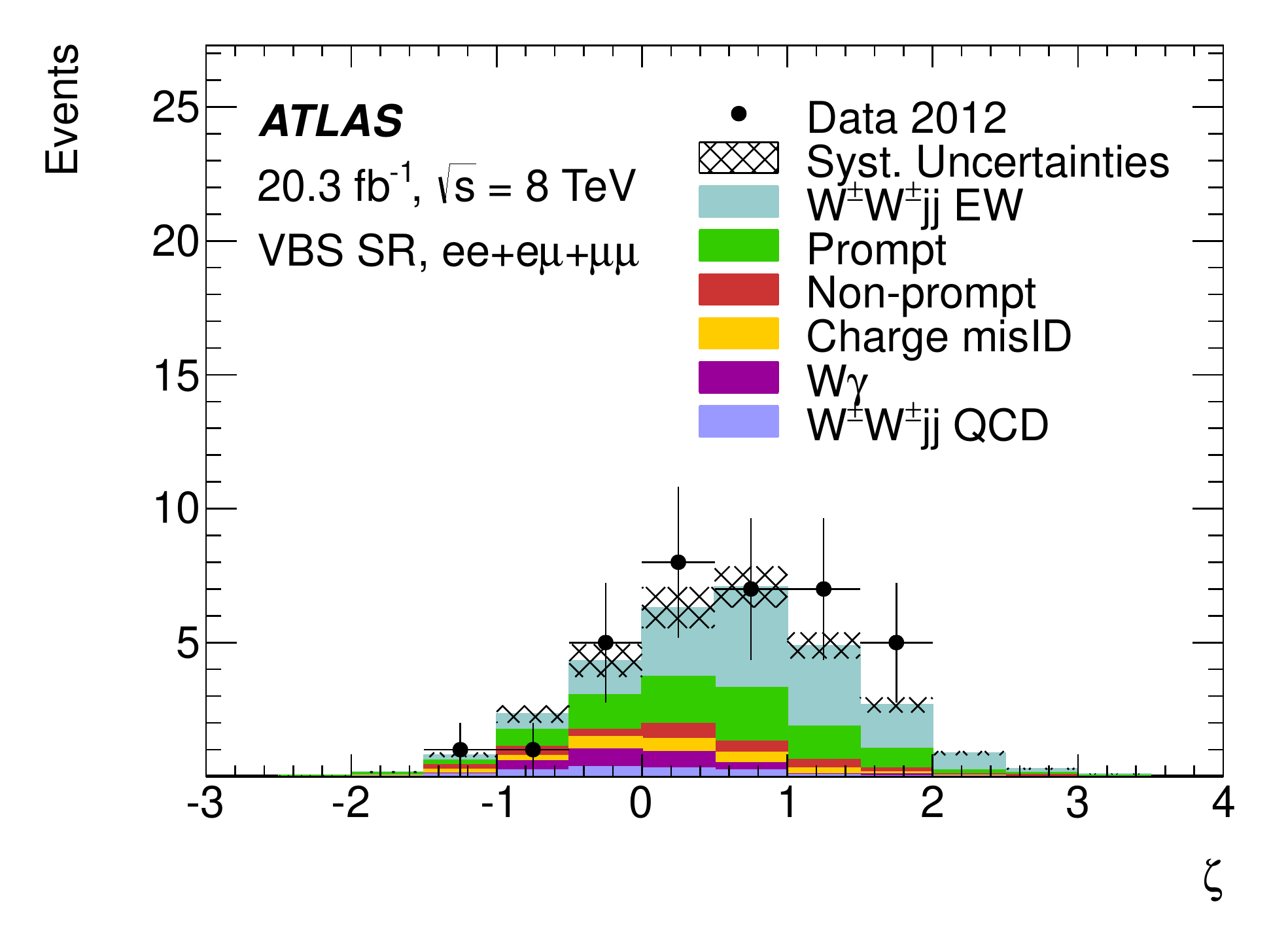}
\includegraphics[width=.48\textwidth]{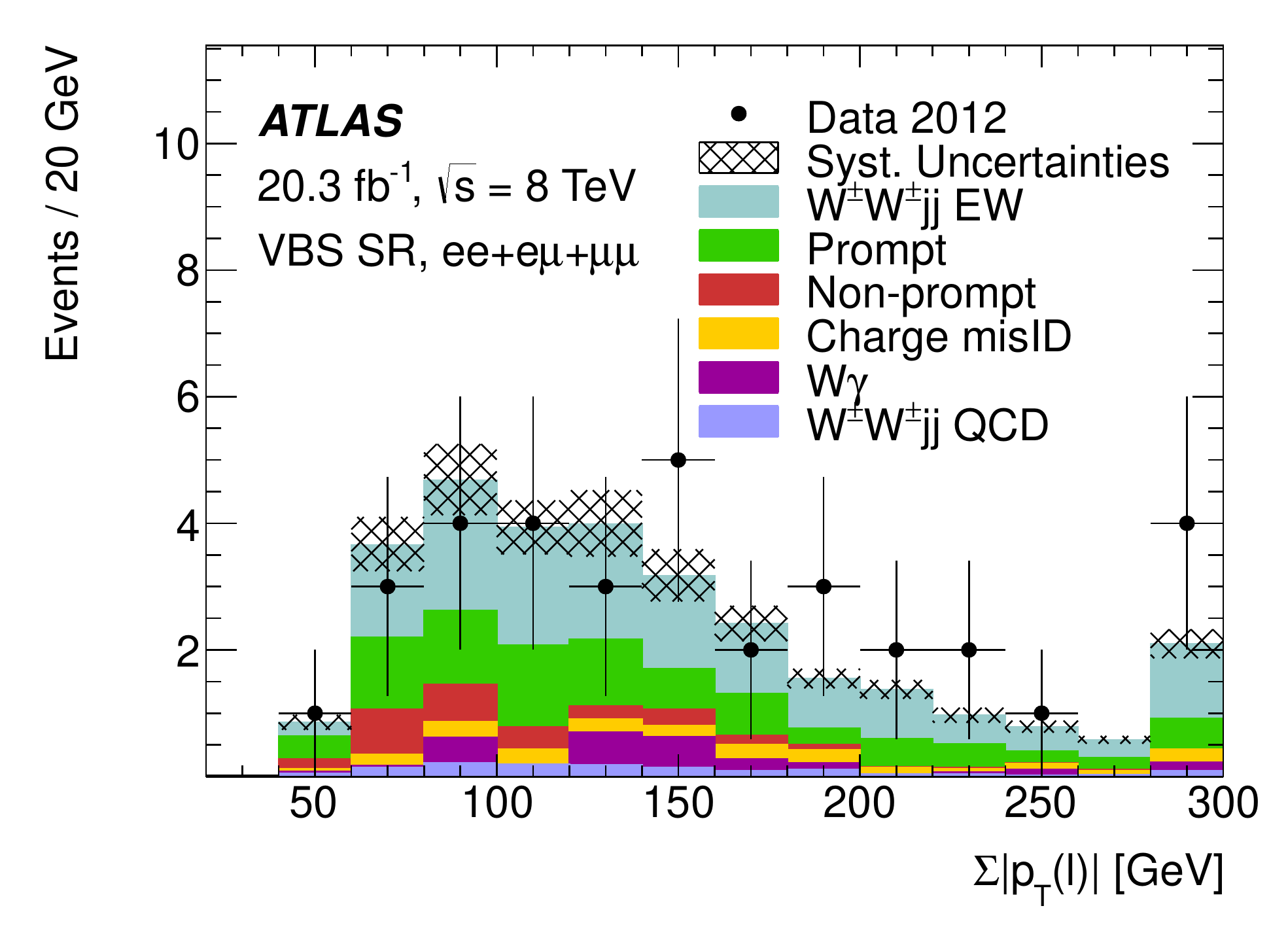}
\caption{The lepton centrality ($\zeta$) distribution (left) and the scalar sum of the two leading leptons' transverse momenta~(right) for all channels combined in the VBS SR. 
The error bars on the data points include statistical uncertainty only. The hatched band represents the systematic uncertainty of the total prediction. The last bin includes overflow events.}
\label{fig:SignalVBS:zetamW}
\end{figure*}

The observed and predicted event yields in the Inclusive and VBS SRs are shown in Tables~\ref{tab:Yields_inclusive} and~\ref{tab:Yields_vbs}, 
broken down by $e^\pm e^\pm$, $e^\pm \mu^\pm$, and $\mu^\pm \mu^\pm$ channels as well as the sum of all three. The observed data events are 
consistent with the SM predictions including \sswwjj\ production. 
Several kinematic distributions are shown in Figures~\ref{fig:SignalVBS:total:mjj}--\ref{fig:SignalVBS:zetamW}. The uncertainties displayed 
are the systematic and statistical uncertainties added in quadrature. All three channels are combined in these plots, and 
correlations of a given systematic uncertainty with others are maintained across signal and background processes and channels. 
The contributions from electroweak and strong \sww production are normalized to the SM predictions. 
Figure~\ref{fig:SignalVBS:total:mjj} presents the dijet invariant mass distribution for the Inclusive SR 
before the final $\mjj > 500~\GeV$ selection is applied.
Figure~\ref{fig:SignalVBS:total:dyjj} presents the \dyjj distribution for the VBS SR before the $\dyjj > 2.4$ selection is applied. 

The lepton centrality is a measure of how central the leptons are with respect to the jets and is defined by
$\zeta = {\textrm min} \left[\eta({\ell_2})-\eta({j_2}), \eta(j_1)-\eta(\ell_1) \right]$, where $\ell_{1,2}$ refers to the two leptons and $j_{1,2}$ refers here to the 
two jets with $\eta(j_1)>\eta(j_2)$, and $\eta(\ell_1)>\eta(\ell_2)$. Events tend to have a lepton centrality greater than zero in the VBS topology. 
The lepton centrality distribution together with the distribution of the scalar sum of the two leading leptons' transverse momenta in the VBS SR are shown in Figure~\ref{fig:SignalVBS:zetamW}. 
Good agreement between data and SM predictions with \sswwjj\ production included is found for all distributions. 

\begin{table*}
\centering
\begin{tabular}{ll|ccc|c}
\hline
\hline
\multicolumn{6}{c}{Inclusive Signal Region}\\ \hline
& & $e^\pm e^\pm$&$e^\pm \mu^\pm$&$\mu^\pm \mu^\pm$&Total \\ \hline
\multicolumn{2}{l|}{\sswwjj-EW} &~~2.82 $\pm$ 0.28&~~7.8 $\pm$ 0.7&4.6 $\pm$ 0.4&15.2 $\pm$ 1.3  \\ \hline
\multicolumn{2}{l|}{\sswwjj-QCD} &~~0.86 $\pm$ 0.15&~~2.3 $\pm$ 0.4&1.45 $\pm$ 0.24&\ \ 4.6 $\pm$ 0.7 \\ \hline
\multicolumn{2}{l|}{Prompt} &~~3.0 $\pm$ 0.7&~~6.1 $\pm$ 1.3&2.6 $\pm$ 0.6&11.6 $\pm$ 2.5 \\ \hline
\multirow{2}{*}{Conversions} & Charge misID &~~2.1 $\pm$ 0.4&~~0.77 $\pm$ 0.27&--&\ \ 2.8 $\pm$ 0.6  \\ \cline{2-6}
& $W\gamma$&~~1.1 $\pm$ 0.6&~~1.6 $\pm$ 0.8&--&\ \ 2.7 $\pm$ 1.2 \\  \hline
\multicolumn{2}{l|}{Non-prompt} &~~0.61 $\pm$ 0.30&~~1.9 $\pm$ 0.8&0.41 $\pm$ 0.22&\ \ 2.9 $\pm$ 0.8  \\ \hline
\multicolumn{2}{l|}{Total predicted} &10.4 $\pm$ 1.3&20.3 $\pm$ 2.5&9.1 $\pm$ 1.0&40 $\pm$ 4  \\ \hline
\multicolumn{2}{l|}{Data} &12&26&12&50 \\
\hline
\hline
\end{tabular}
\caption{Predicted and observed numbers of events in the inclusive SR are shown separately for the \elel, \elmu, 
and \mum channels as well as for the sum of all three. 
The uncertainty is the combination of statistical and systematic uncertainties; correlations among systematic uncertainties are taken into account in the calculations of the total. 
The contributions from \sswwjj-EW and \sswwjj-QCD production are normalized to the SM prediction.} 
\label{tab:Yields_inclusive}
\end{table*} 

\begin{table*}
\centering
\begin{tabular}{ll|ccc|c}
\hline
\hline
\multicolumn{6}{c}{VBS Signal Region}\\
\hline
& & $e^\pm e^\pm$&$e^\pm \mu^\pm$&$\mu^\pm \mu^\pm$&Total\\
\hline
 \multicolumn{2}{l|}{\sswwjj-EW} &2.34 $\pm$ 0.23&~~6.3 $\pm$ 0.6&3.77 $\pm$ 0.35&12.4 $\pm$ 1.1\\ \hline
 \multicolumn{2}{l|}{\sswwjj-QCD} &0.26 $\pm$ 0.06&~~0.67 $\pm$ 0.14&0.43 $\pm$ 0.09&\ \ 1.36 $\pm$ 0.27\\ \hline
 \multicolumn{2}{l|}{Prompt} &2.2 $\pm$ 0.5&~~4.2 $\pm$ 1.0&1.9 $\pm$ 0.5&\ \ 8.2 $\pm$ 1.9\\ \hline
 \multirow{2}{*}{Conversions}  & Charge misID &1.39 $\pm$ 0.27&~~0.64 $\pm$ 0.24&--&\ \ 2.0 $\pm$ 0.5\\ \cline{2-6}
& $W\gamma$&0.7 $\pm$ 0.4&~~1.3 $\pm$ 0.7& -- &\ \ 2.0 $\pm$ 1.0\\   \hline
 \multicolumn{2}{l|}{Non-prompt} &0.50 $\pm$ 0.26&~~1.5 $\pm$ 0.6&0.34 $\pm$ 0.19&\ \ 2.3 $\pm$ 0.7\\ \hline
 \multicolumn{2}{l|}{Total predicted} &7.4 $\pm$ 1.0&14.5 $\pm$ 1.9&6.4 $\pm$ 0.7&28.3 $\pm$ 3.4\\ \hline
 \multicolumn{2}{l|}{Data} & 6&18&10&34\\
\hline
\hline
\end{tabular}
\caption{Predicted and observed numbers of events in the VBS SR are shown separately for the \elel, \elmu, 
and \mum channels as well as for the sum of all three. 
The uncertainty is the combination of statistical and systematic uncertainties; correlations among systematic uncertainties are taken into account in the calculations of the total. 
The contributions from \sswwjj-EW and \sswwjj-QCD production are normalized to the SM prediction. } 
\label{tab:Yields_vbs}
\end{table*} 

The data are also divided into \ensuremath{W^{+}W^{+}} and \ensuremath{W^{-}W^{-}} channels. The \ensuremath{W^{+}W^{+}} channel is favored by data and SM prediction 
as the LHC is a $pp$ collider. These two channels are not split by leptonic final states due to the limited number of events. 
The event yields are shown in Table~\ref{tab:IncYields}, and the observed charge distribution in data is found to be consistent with SM predictions. 

\begin{table}
\centering
\begin{tabular}{ll|cc|cc}
\hline
\hline
& &\multicolumn{2}{c|}{Inclusive Signal Region} & \multicolumn{2}{c}{VBS Signal Region}\\
& &$W^{+} W^{+}$&$W^{-} W^{-}$ &$W^{+} W^{+}$&$W^{-} W^{-}$\\
\hline
\multicolumn{2}{l|}{\sswwjj-EW}  &13.0 $\pm$ 1.2&\ \ 3.9 $\pm$ 0.4 &\ \ 9.4 $\pm$ 0.8&2.90 $\pm$ 0.27\\ \hline
\multicolumn{2}{l|}{\sswwjj-QCD} &\ \ 3.6 $\pm$ 0.6&\ \ 1.14 $\pm$ 0.19 &\ \ 1.08 $\pm$ 0.21&0.26 $\pm$ 0.06\\ \hline
\multicolumn{2}{l|}{Prompt} &\ \ 8.0 $\pm$ 1.7&\ \ 3.7 $\pm$ 0.8 &\ \ 6.0 $\pm$ 1.4&2.2 $\pm$ 0.6\\ \hline
\multirow{2}{*}{Conversions} & Charge misID &\ \ 1.27 $\pm$ 0.28&\ \ 1.57 $\pm$ 0.35 &\ \ 0.90 $\pm$ 0.23&1.13 $\pm$ 0.28\\ \cline{2-6}
& $W\gamma$&\ \ 1.7 $\pm$ 0.8&\ \ 1.0 $\pm$ 0.6 &\ \ 1.4 $\pm$ 0.7&0.6 $\pm$ 0.4\\  \hline
\multicolumn{2}{l|}{Non-prompt}&\ \ 1.7 $\pm$ 0.5&\ \ 1.2 $\pm$ 0.4 &\ \ 1.4 $\pm$ 0.4&0.95 $\pm$ 0.33\\ \hline
\multicolumn{2}{l|}{Total predicted}&29.3 $\pm$ 3.3&12.5 $\pm$ 1.6 & 20.2 $\pm$ 2.5&8.1 $\pm$ 1.4\\ \hline
\multicolumn{2}{l|}{Data}&35&15 & 23&11\\ \hline
\hline
\end{tabular}
\caption{Event yields for predicted signal and background events as well as observed data in the VBS SR for the \ensuremath{W^{+}W^{+}} and \ensuremath{W^{-}W^{-}} channels.
The uncertainty is the combination of statistical and systematic uncertainties; correlations among systematic uncertainties are taken into account in the calculations of the total. 
}
\label{tab:IncYields}
\end{table}

\section{Extraction of production cross sections}
The excesses in data over the background-only predictions in the Inclusive and VBS SRs are consistent with the event topology for \sswwjj~production. 
The numbers of observed data and expected signal and background events are used to calculate the fiducial cross-sections in these two signal regions.  

\noindent\textbf{Cross-section extraction method} \\
A likelihood function is used to extract the cross-sections in the two fiducial regions. 
The likelihood function uses Poisson distributions for each channel and global constraints for the nuisance 
parameters $\theta_j$, which parameterize effects of systematic uncertainties.
The number of expected events in a given decay channel $c$, $N^{\text{exp}}_c$, is a product of the integrated luminosity \lum, the measured fiducial cross-section
$\sigma_{W^\pm W^\pm jj}$, the relative acceptance for each channel, $A_c$, and the signal efficiency 
$\epsilon_c$, in addition to the total number of background events in this channel, $\sum_{b} N_{c,b}$:

\begin{equation}
N^{\text{exp}}_c= \lum \cdot \sigma_{W^\pm W^\pm jj} \cdot A_c \cdot \varepsilon_c+ \sum_{b} N_{c,b} ~. \label{eq:Nexp} 
\end{equation}

The likelihood function is given by

\begin{equation}
L = \prod_c \text{Pois}\left(N^\text{obs}_c|N_c^\text{exp}\right) \prod_j g \left(0 | \theta_j,1 \right) ~.
\label{eq:llh}
\end{equation}

The function $g$ is a Gaussian probability density function.  
The effect due to systematic uncertainties in $\varepsilon_c$ and $N_{c,b}$ are parameterized by the nuisance parameters according to

\begin{align}
\varepsilon_c(\theta_j) &= \varepsilon_c^0 \prod_j \left( 1 + \theta_j \delta_{c,j}^s \right) \ ,\\
N_{c,b}(\theta_j) &= N_{c,b}^0 \prod_j \left( 1 + \theta_j \delta_{c,j}^b \right)
\end{align}

with $\varepsilon_c^0$ and $N_{c,b}^0$ being the nominal estimates for the signal reconstruction efficiency and the background yields in channel~$c$. 
The constants $\delta_{c,j}^{s}$ and $\delta_{c,j}^{b}$ represent the relative uncertainty in the signal reconstruction efficiency and the nominal 
background prediction, respectively, in channel~$c$ due to the source of systematic uncertainty,~$j$.

The relative acceptances within the fiducial region are determined at
particle level from the decay
branching ratios of the two $W$ bosons to 
$e^\pm e^\pm$, $e^\pm \mu^\pm$, and $\mu^\pm \mu^\pm$. 
Small deviations arise from the jet object definition at particle level, which accepts electrons as input objects to the jet clustering 
algorithm while muons are ignored. The acceptances in the corresponding channels are 0.232, 0.524, and 0.265 in the Inclusive SR 
and 0.235, 0.527, and 0.257 in the VBS SR, respectively.

The signal efficiency for channel $c$, $\varepsilon_c$, is estimated from simulated signal events. It is given by the number of 
events reconstructed in a given signal region divided by the number of events
passing the corresponding definition of the fiducial phase-space region at the particle level. 
It accounts for the detector reconstruction, particle identification, and trigger efficiency as well as for the migration into and out of the fiducial volume due 
to detector resolution effects. The signal efficiency definition includes contributions from leptons originating from $\tau$ decays at the reconstruction level, while 
those events are vetoed at the particle level.  
The fraction of events where the electron or muon originates from a $\tau$ lepton in the signal yield at the reconstruction level is found to be 10\%. 
The efficiencies in the $e^\pm e^\pm$, $e^\pm \mu^\pm$, and $\mu^\pm \mu^\pm$ channels are $(56.2 \pm 1.5)\%$,  $(71.7 \pm 0.8)\%$, 
and $(77.0 \pm 0.9)\%$ in the Inclusive signal region and $(57.2 \pm 1.6)\%$, $(72.7 \pm 1.0)\%$, and $(82.7 \pm 1.2)\%$ in the VBS signal region, respectively.

The measured cross-sections are taken as those maximizing the log-likelihood function shown in Eq.~\eqref{eq:llh}. 
The quoted uncertainties are derived using the profile likelihood method~\cite{profile} and correspond to likelihood intervals with a confidence level (CL) of 68.3\%.

\begin{figure*}[pht!]
\centering
\includegraphics[width=0.48\textwidth]{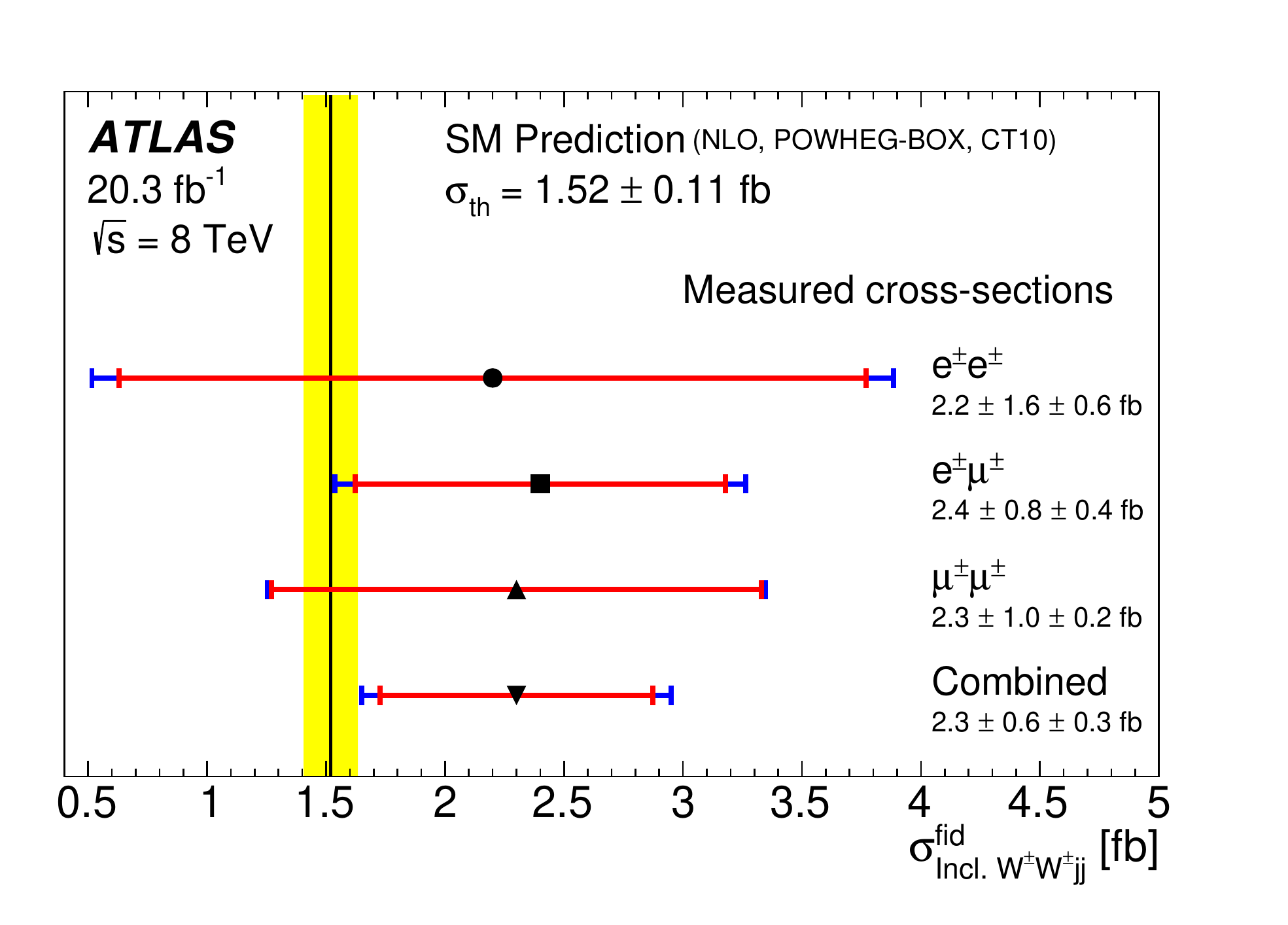}
\includegraphics[width=0.48\textwidth]{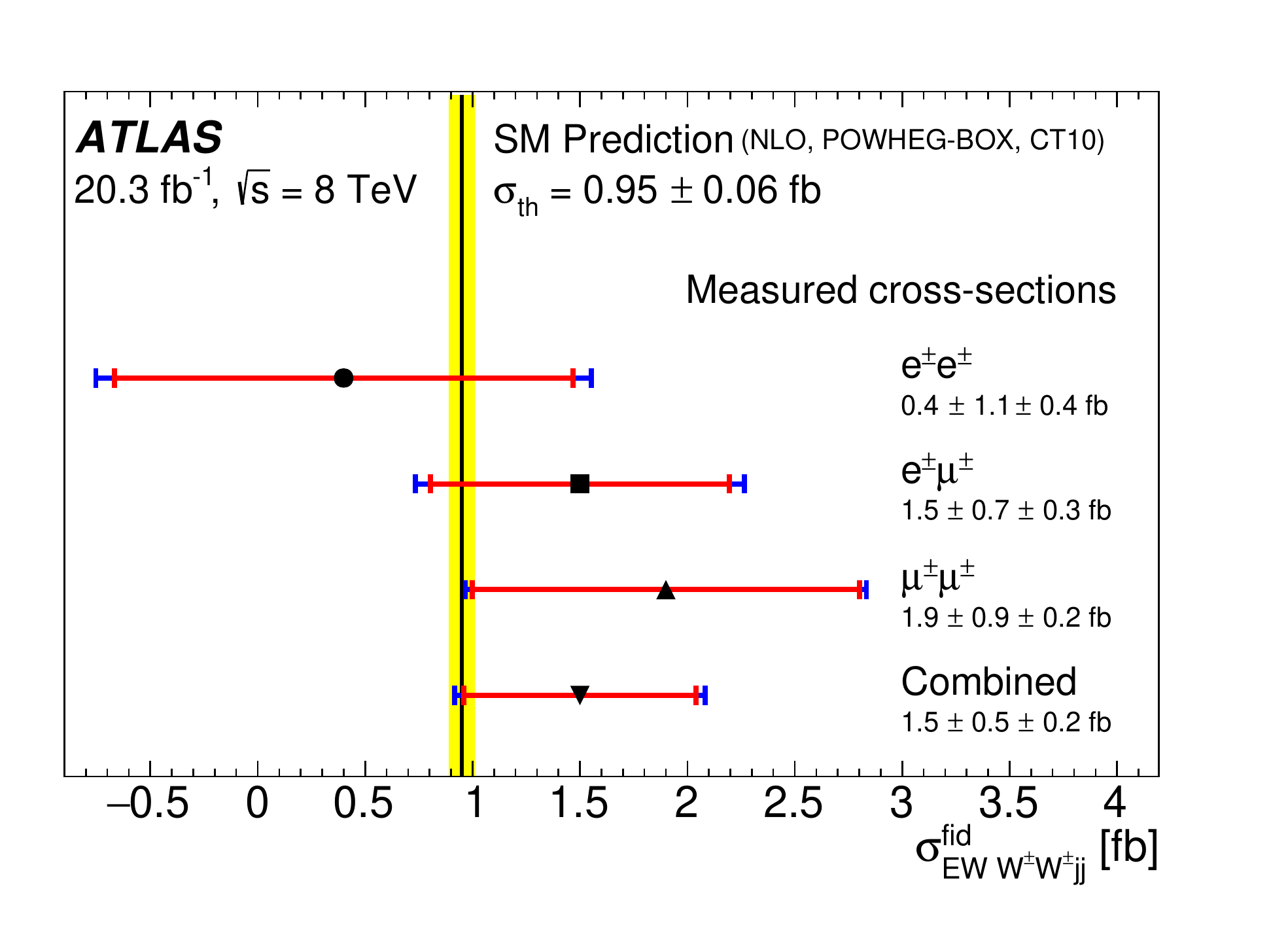}
\caption{The measured cross-sections for the Inclusive SR (left) and the VBS SR (right) compared to the 
predictions for each channel and for the combined measurement. The inner error band represents the statistical uncertainty and the outer band represents the total uncertainty of each measurement.}
\label{fig:xsec_mes} 
\end{figure*}

\bigskip 
\noindent\textbf{Measured fiducial cross-sections} \\
The measured fiducial cross-section is 
\mbox{$\sigma_{{\textrm Incl.}~W^\pm W^\pm jj}^{\textrm fid}=2.3\pm0.6 ({\text{stat}})\pm0.3({\text{syst}})$~fb} for the \sswwjj\ production, including both electroweak and 
strong production as well as the interference in the Inclusive SR. 
The measured fiducial cross-section is \mbox{$\sigma_{{\textrm EW}~W^\pm W^\pm jj}^{\textrm fid}=1.5\pm0.5({\text{stat}})\pm0.2({\text{syst}})$}~fb 
for electroweak \sww production, including interference with strong production in the VBS region.
The measured cross-sections are in agreement with the respective SM predictions of $1.52\pm0.11$~fb and $0.95\pm0.06$~fb.
The cross-sections are shown in Figure~\ref{fig:xsec_mes} for each channel and for the combined measurement.
The observed combined significance over the background-only hypothesis is 4.5$\sigma$ 
in the Inclusive SR and 3.6$\sigma$ in the VBS SR, while the corresponding expected significances for a SM \sswwjj\ signal are 3.1$\sigma$ and 2.3$\sigma$, respectively.

\section{Extraction of anomalous quartic gauge couplings}

VBS events receive contributions from quartic gauge boson interactions and thus can be used to search for aQGCs. 
In general, the effective Lagrangian described in Section~\ref{sec:introduction} does not ensure unitarity. The Higgs boson in the SM ensures unitarity of the SM VBS process, which is destroyed if
anomalous couplings or additional resonances are added. A unitarization scheme has to be applied in order to avoid non-physical
predictions. In the case of VBS with aQGC, the unitarization significantly impacts the
differential and total cross-sections. 
The K-matrix unitarization scheme~\cite{kmatrix} is applied in this analysis where the elastic scattering
eigen-amplitude~$\mathcal{A}(s)$ is projected on the Argand circle
${\mathcal{A}(s) \rightarrow \hat{\mathcal{A}}(s)}$ such that
$|\hat{\mathcal{A}}(s) - \mathrm{i}/2| = 1/2$. This condition is derived
from the optical theorem and ensures that the projected scattering amplitude
meets the unitarity condition exactly. As a result, the cross-section saturates at the
maximum value allowed by unitarity. The {\textsc whizard}~\cite{whizard1} event generator is 
used to calculate cross-sections and generate events with aQGCs at LO in QCD. 
The {\textsc CTEQ6L1} PDF set is used. All samples use the parameterization in terms
of $\alpha_4$ and $\alpha_5$. The invariant mass of the system of two charged leptons and two neutrinos from the 
decay of the two $W$ bosons, $m_{\ell\ell\nu\nu}$, is used as the
renormalization and factorization scales, $\mu_\mathrm{R}=\mu_\mathrm{F}=m_{\ell\ell\nu\nu}$. The
events are interfaced to {\textsc pythia} 8 for modeling the parton shower, QED final-state radiation, 
decays of $\tau$ leptons, and the underlying event.

\begin{figure}
  \centering
  \includegraphics[width=0.48\linewidth]{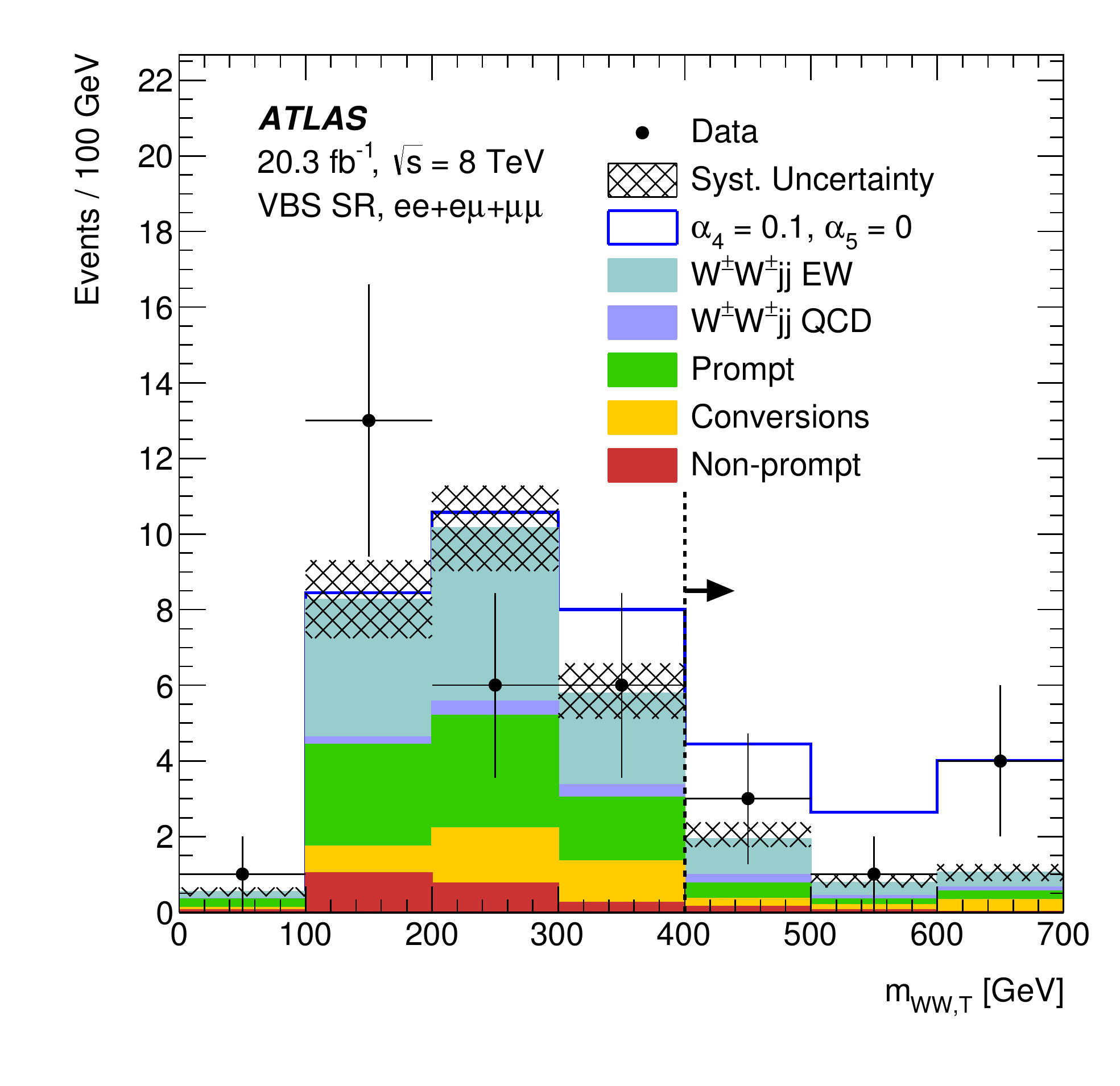}
  \caption{The $m_{WW,\mathrm{T}}$ distribution for all channels combined in the VBS SR prior to applying the requirement of $m_{WW, \mathrm{T}}>400$ \GeV. The $m_{WW, \mathrm{T}}$ requirement is represented by a vertically dashed line. The expected signal contribution for the aQGC parameter point $\alpha_4 = 0.1$ and $\alpha_5 = 0$ is overlaid as a histogram and includes the aQGC signal and the background prediction. The error bars on the data points include statistical uncertainty only. The hatched band represents the systematic uncertainty of the total prediction. The last bin includes overflow events.}
  \label{fig:VBS_mvec}
\end{figure}

The expected sensitivity to $\alpha_4$ and $\alpha_5$ is improved significantly compared to the results obtained in 
the previous publication~\cite{ssWWPRL} by selecting a phase-space region that is more sensitive to anomalous contributions
to the $WWWW$ vertex. This is achieved by an additional requirement: $m_{WW,\mathrm{T}}$ > 400 \GeV. The effects from new-physics processes are
expected to be seen predominantly at larger mass scales, which motivates the definition of the aQGC SR as defined in Section~\ref{sec:Analysis_Selection}. 
The distribution of the transverse mass of the $WW$ system before applying the final selection criteria is shown in Figure~\ref{fig:VBS_mvec}.

The signal in the aQGC region is defined as the $\alpha_4,\alpha_5$-dependent excess of the \sswwjj-EW production cross-section over the SM
prediction of this process. No interference effects of the aQGC contribution with either the SM \sswwjj-QCD or \sswwjj-EW production are considered. 
The combined signal reconstruction efficiency in the three final states is found to be $(68.7 \pm 2.2)\%$ with no significant
dependence on $\alpha_4$ and $\alpha_5$.

Table~\ref{tab:event_yields_aqgc} summarizes the expected and observed event
yields in the aQGC SR. The theoretical uncertainties in the aQGC signal region
are less than in the VBS region and the systematic uncertainties are consistent with those in the VBS signal region. 
Therefore, the VBS signal region systematic uncertainties as described in Section~\ref{sec:systematics} are applied. A total of ${3.8 \pm 0.6}$ events are expected from {SM} background processes. 
The expected number of additional events for the {aQGC}
parameter point ${\alpha_4 = 0.1}$ and ${\alpha_5 = 0}$ is also shown. 
In total 8 events are observed in data, which corresponds to an excess with a significance of 1.8$\sigma$.

\begin{table}
 \centering
 \begin{tabular}{lr@{ $\pm$ }r@{ $\pm$ }l}
 \hline
 \hline
             & \multicolumn{3}{c}{aQGC Signal Region} \\
 \hline
Non-prompt  & 0.2 & 0.1 & 0.1 \\
Conversions & 0.7 & 0.2 & 0.1 \\
Prompt     & 0.8 & 0.1 & 0.3 \\
SM \WWssjj{}-EW        & 1.7 & 0.1 & 0.2 \\
SM \WWssjj{}-QCD        & 0.4 & 0.0 & 0.1 \\
\hline
Total background       & 3.8 & 0.3 & 0.5 \\
\hline
$\alpha_4 = 0.1$, $\alpha_5 = 0$ & 7.3 & 0.4 & 0.6 \\
\hline
Data & \multicolumn{3}{l}{8} \\
\hline
\hline
\end{tabular}
 \caption[Event yields in the aQGC signal region]{Expected and observed event yields in the aQGC SR. 
The first quoted uncertainty is statistical and the second is systematic. The row corresponding to the BSM contribution indicates the additional events expected given $\alpha_4 = 0.1$ and $\alpha_5 = 0$.}
\label{tab:event_yields_aqgc}
\end{table}

A CL$_{s}$ upper limit~\cite{CLs} on the visible cross-section in the aQGC SR 
is reported. The visible cross-section~$\sigma^\text{vis}$ is defined
at the detector level as the excess of data events ($N^\text{obs}$) over the background prediction
($N^\text{bkg}$) divided by the integrated luminosity:

\begin{equation}
  \sigma^\text{vis} = \frac{N^\text{obs} - N^\text{bkg}}{\lum} \ .
\end{equation}

The CL$_s$ upper limit is derived with a likelihood function
equivalent to the one defined in Eq.~\eqref{eq:llh} for a single
channel by replacing $\sigma_{W^\pm W^\pm jj} \cdot A_c
\cdot \varepsilon_c$ with $\sigma^\text{vis}$ in
Eq.~\eqref{eq:Nexp} where $\sigma^\text{vis}$ is 
affected by uncertainties in the background prediction and the integrated luminosity, 
but not by reconstruction efficiencies or uncertainties in the theoretical cross-sections of the SM $W^\pm W^\pm jj$ production.
The observed (expected) 95\% CL upper limit on $\sigma^\text{vis}$ in the aQGC SR is 0.50~fb (0.25~fb). 
These limits are converted to upper limits on the fiducial cross-section, assuming the same signal reconstruction efficiency as that of the \sswwjj-EW production. 
Models predicting contributions to the aQGC fiducial phase-space region at the 
particle level of more than 0.72~fb (0.37~fb) are excluded at the 95\% CL. 

\begin{figure}[h!]
 \centering
\includegraphics[width=0.55\linewidth]{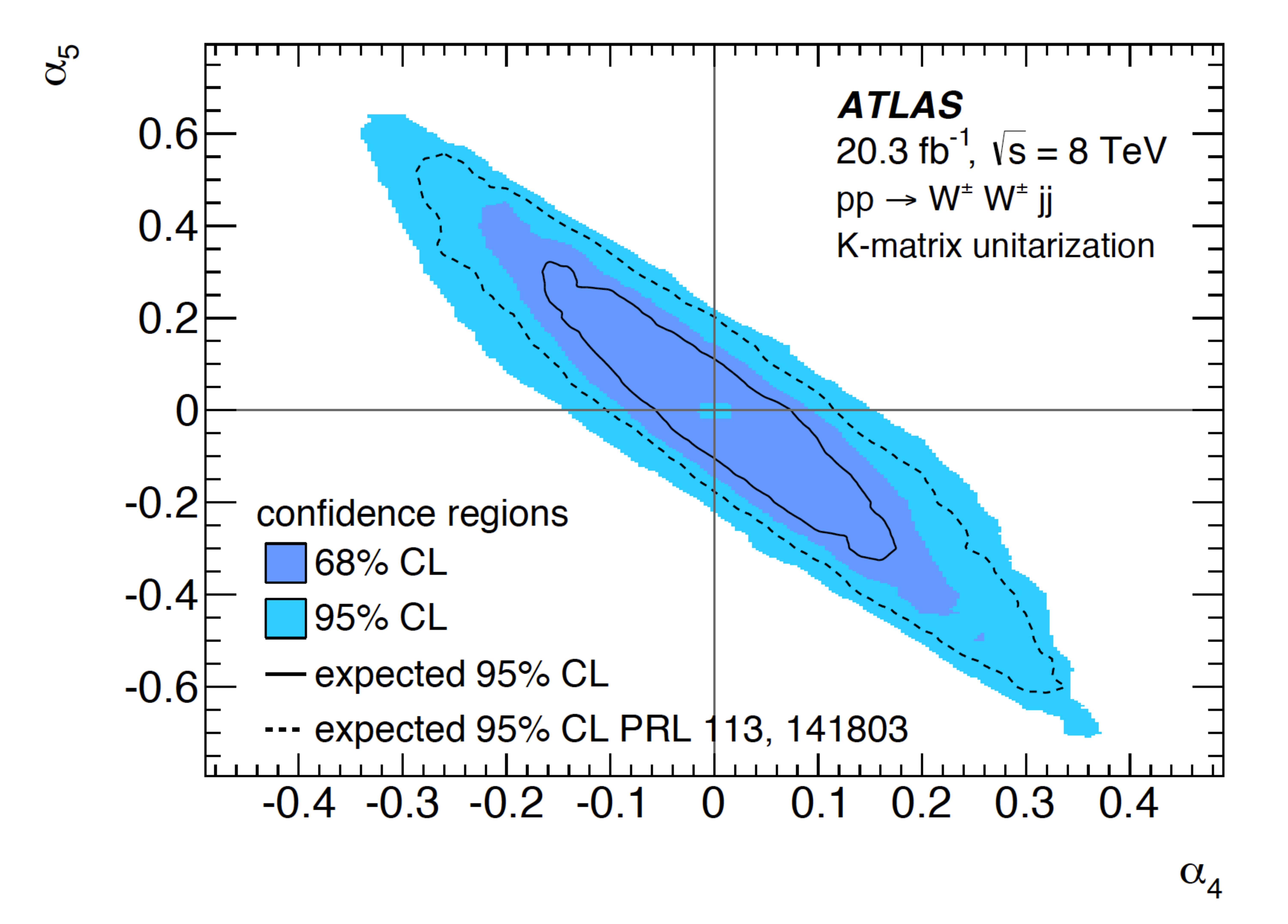}
 \caption[Two-dimensional confidence intervals on the {aQGC} parameters]{Two-dimensional confidence regions in the {aQGC} parameter plane ($\alpha_4,\alpha_5$). 
The area outside the solid light blue region is excluded by the data at the 95\% CL. The area outside the solid dark blue region is 
excluded at the 68\% CL. The expected exclusion contour at the 95\% CL is marked by the solid black line. For comparison, 
the expected exclusion contour at the 95\% CL from the previous analysis of this final state~\cite{ssWWPRL} is shown as a black dashed line.}
 \label{fig:aQGC_limits_aqgc}
\end{figure}

The upper limits on the fiducial cross-section in the aQGC phase-space region at the particle level are used to derive constraints in the ($\alpha_4, ~\alpha_5$) parameter
space. The expected and observed two-dimensional exclusion contours
are shown in Figure~\ref{fig:aQGC_limits_aqgc}. The expected one-dimensional
confidence intervals at the 95\% CL are

\begin{equation*}
  \alpha_4 \in [-0.06,0.07], \;\; {\textrm and} \; \; \alpha_5 \in [-0.10,0.11] \; \; {\textrm(expected)}\ .
\end{equation*}

The observed one-dimensional confidence intervals at the 95\% CL are 

\begin{equation*}
  \alpha_4 \in [-0.14,0.15], \;\; {\textrm and} \; \; \alpha_5 \in [-0.22,0.22] \; \; {\textrm(observed)}\ .
\end{equation*}

This result constitutes a 35\% improvement in the expected aQGC sensitivity with
respect to the analysis published in Ref.~\cite{ssWWPRL}. The observed exclusion is only marginally more restrictive because of the small excess
observed in the aQGC signal region. The sensitivity is similar to that in Ref.~\cite{WW-WZ}, where the observed results are more constraining.

\section{Summary}
This paper presents results from the ATLAS detector at the LHC using 20.3 fb$^{-1}$ of proton--proton collision data at $\sqrt{s}$ = 8 TeV from the measurement of the \sswwjj\ production cross-sections. 
Events with two leptons (electrons or muons) with the same electric charge, \met, and at least two jets are investigated in the Inclusive signal region. 
An additional selection on the rapidity difference of the leading jets is used to measure the fiducial cross-section for 
the \sswwjj-EW production in the VBS signal region. The further requirement of a high transverse mass of the system of two leptons 
and \met\ is used to define a restricted phase-space region more sensitive to aQGC parameters.

In the Inclusive signal region, a total of 50 signal candidates are observed and 20 background events are expected. 
The excess of events over the background-only prediction is interpreted as evidence for the sum of the \sswwjj-EW and \sswwjj-QCD processes.
The measured fiducial cross-section for \sswwjj\ production is
$2.3\pm0.6\mbox{(stat.)} \pm0.3\mbox{(syst.)} $ fb, with a significance of 4.5$\sigma$ (3.1$\sigma$ expected).
In the VBS signal region, the background-only prediction includes the \sswwjj-QCD production, and
a total of 34 events are observed and 16 background events are predicted. 
The excess is interpreted as evidence for the \sswwjj-EW processes. The measured 
fiducial cross-section for the \sswwjj-EW production, including the interference with the \sswwjj-QCD production, is
$1.5\pm0.5\mbox{(stat.)} \pm0.2\mbox{(syst.)} $ fb 
with a significance of 3.6$\sigma$ (2.3$\sigma$ expected).
The measured cross-sections are consistent with the SM predictions. 

In the aQGC signal region, the background prediction includes both the \sswwjj-EW and \sswwjj-QCD processes. A total of 8 events are observed and
3.8 background events are expected. These numbers are used to constrain the aQGC parameters $\alpha_{4}$ and $\alpha_{5}$.
The observed one-dimensional 95\% confidence level intervals are $-0.14 < \alpha_4 < 0.15$ and $-0.22 < \alpha_5 < 0.22$. 
The expected 95\% confidence level intervals are $-0.06 < \alpha_4 < 0.07$ and $-0.10 < \alpha_5 < 0.11$. 
These intervals constitute a 35\% improvement in the expected aQGC sensitivity with respect to the analysis published in Ref.~\cite{ssWWPRL}.

\section*{Acknowledgments}
We thank CERN for the very successful operation of the LHC, as well as the
support staff from our institutions without whom ATLAS could not be
operated efficiently.

We acknowledge the support of ANPCyT, Argentina; YerPhI, Armenia; ARC, Australia; BMWFW and FWF, Austria; ANAS, Azerbaijan; SSTC, Belarus; CNPq and FAPESP, Brazil; NSERC, NRC and CFI, Canada; CERN; CONICYT, Chile; CAS, MOST and NSFC, China; COLCIENCIAS, Colombia; MSMT CR, MPO CR and VSC CR, Czech Republic; DNRF and DNSRC, Denmark; IN2P3-CNRS, CEA-DSM/IRFU, France; SRNSF, Georgia; BMBF, HGF, and MPG, Germany; GSRT, Greece; RGC, Hong Kong SAR, China; ISF, I-CORE and Benoziyo Center, Israel; INFN, Italy; MEXT and JSPS, Japan; CNRST, Morocco; NWO, Netherlands; RCN, Norway; MNiSW and NCN, Poland; FCT, Portugal; MNE/IFA, Romania; MES of Russia and NRC KI, Russian Federation; JINR; MESTD, Serbia; MSSR, Slovakia; ARRS and MIZ\v{S}, Slovenia; DST/NRF, South Africa; MINECO, Spain; SRC and Wallenberg Foundation, Sweden; SERI, SNSF and Cantons of Bern and Geneva, Switzerland; MOST, Taiwan; TAEK, Turkey; STFC, United Kingdom; DOE and NSF, United States of America. In addition, individual groups and members have received support from BCKDF, the Canada Council, CANARIE, CRC, Compute Canada, FQRNT, and the Ontario Innovation Trust, Canada; EPLANET, ERC, ERDF, FP7, Horizon 2020 and Marie Sk{\l}odowska-Curie Actions, European Union; Investissements d'Avenir Labex and Idex, ANR, R{\'e}gion Auvergne and Fondation Partager le Savoir, France; DFG and AvH Foundation, Germany; Herakleitos, Thales and Aristeia programmes co-financed by EU-ESF and the Greek NSRF; BSF, GIF and Minerva, Israel; BRF, Norway; CERCA Programme Generalitat de Catalunya, Generalitat Valenciana, Spain; the Royal Society and Leverhulme Trust, United Kingdom.

The crucial computing support from all WLCG partners is acknowledged gratefully, in particular from CERN, the ATLAS Tier-1 facilities at TRIUMF (Canada), NDGF (Denmark, Norway, Sweden), CC-IN2P3 (France), KIT/GridKA (Germany), INFN-CNAF (Italy), NL-T1 (Netherlands), PIC (Spain), ASGC (Taiwan), RAL (UK) and BNL (USA), the Tier-2 facilities worldwide and large non-WLCG resource providers. Major contributors of computing resources are listed in Ref.~\cite{ATL-GEN-PUB-2016-002}.

\printbibliography

\newpage \input{atlas_authlist}

\end{document}

%% file: atlas_authlist.tex
\begin{flushleft}
{\Large The ATLAS Collaboration}

\bigskip

M.~Aaboud$^\textrm{\scriptsize 136d}$,
G.~Aad$^\textrm{\scriptsize 87}$,
B.~Abbott$^\textrm{\scriptsize 114}$,
J.~Abdallah$^\textrm{\scriptsize 65}$,
O.~Abdinov$^\textrm{\scriptsize 12}$,
B.~Abeloos$^\textrm{\scriptsize 118}$,
R.~Aben$^\textrm{\scriptsize 108}$,
O.S.~AbouZeid$^\textrm{\scriptsize 138}$,
N.L.~Abraham$^\textrm{\scriptsize 152}$,
H.~Abramowicz$^\textrm{\scriptsize 156}$
% ...
\bigskip
\\
% ...
$^{12}$ Institute of Physics, Azerbaijan Academy of Sciences, Baku, Azerbaijan\\
% ...
$^{65}$ University of Iowa, Iowa City IA, United States of America\\
% ...
$^{87}$ CPPM, Aix-Marseille Universit{\'e} and CNRS/IN2P3, Marseille, France\\
% ...
$^{108}$ Nikhef National Institute for Subatomic Physics and University of Amsterdam, Amsterdam, Netherlands\\
% ...
$^{114}$ Homer L. Dodge Department of Physics and Astronomy, University of Oklahoma, Norman OK, United States of America\\
% ...
$^{118}$ LAL, Univ. Paris-Sud, CNRS/IN2P3, Universit{\'e} Paris-Saclay, Orsay, France\\
% ...
$^{136}$ $^{(a)}$ Facult{\'e} des Sciences Ain Chock, R{\'e}seau Universitaire de Physique des Hautes Energies - Universit{\'e} Hassan II, Casablanca; $^{(b)}$ Centre National de l'Energie des Sciences Techniques Nucleaires, Rabat; $^{(c)}$ Facult{\'e} des Sciences Semlalia, Universit{\'e} Cadi Ayyad, LPHEA-Marrakech; $^{(d)}$ Facult{\'e} des Sciences, Universit{\'e} Mohamed Premier and LPTPM, Oujda; $^{(e)}$ Facult{\'e} des sciences, Universit{\'e} Mohammed V, Rabat, Morocco\\
% ...
$^{138}$ Santa Cruz Institute for Particle Physics, University of California Santa Cruz, Santa Cruz CA, United States of America\\
% ...
$^{152}$ Department of Physics and Astronomy, University of Sussex, Brighton, United Kingdom\\
% ...
$^{156}$ Raymond and Beverly Sackler School of Physics and Astronomy, Tel Aviv University, Tel Aviv, Israel\\
% ...
$^{a}$ Also at Department of Physics, King's College London, London, United Kingdom\\
$^{b}$ Also at Institute of Physics, Azerbaijan Academy of Sciences, Baku, Azerbaijan\\
$^{c}$ Also at Novosibirsk State University, Novosibirsk, Russia\\
$^{d}$ Also at TRIUMF, Vancouver BC, Canada\\
$^{e}$ Also at Department of Physics {\&} Astronomy, University of Louisville, Louisville, KY, United States of America\\
% ...
\end{flushleft}
